\documentclass[traditabstract]{aa}
\usepackage{graphicx}
\usepackage{txfonts}
\usepackage{color}
\usepackage{txfonts}
\usepackage[]{natbib}
\bibpunct{(}{)}{;}{a}{}{,}
\makeatletter 
\renewcommand\@biblabel[1]{} 
\makeatother %

\usepackage{epsfig}
\usepackage{graphics}
\usepackage{float}
\usepackage{multirow}
\usepackage{longtable}
\usepackage{rotate}
\usepackage{amssymb}
\usepackage{array}
\usepackage{times}
\usepackage{longtable}

\def\kms{\ifmmode{\rm km\,s^{-1}}\else\hbox{$\rm km\,s^{-1}$}\fi}

\begin{document}
\title{Extensive optical and near-infrared observations of the nearby, narrow-lined type Ic \object{SN~2007gr}:
       days 5 to 415}
\author{D. J. Hunter \inst{1}, S. Valenti \inst{1}, R. Kotak \inst{1}, W. P. S. Meikle \inst{2}, S. Taubenberger \inst{3}, A. Pastorello \inst{1}, S. Benetti \inst{4}, V. Stanishev \inst{4},  S. J. Smartt \inst{1}, C. Trundle \inst{1}, A. A. Arkharov \inst{5, 6}, M. Bufano \inst{7}, E. Cappellaro \inst{7}, E. Di Carlo \inst{7}, M. Dolci \inst{7}, N. Elias-Rosa \inst{8}, S. Frandsen \inst{9}, J. U. Fynbo \inst{10}, U. Hopp \inst{11, 12}, V. M. Larionov \inst{5,6}, P. Laursen \inst{10}, P. Mazzali \inst{3, 4, 13}, H. Navasardyan \inst{4}, C. Ries \inst{11}, A. Riffeser \inst{11}, L. Rizzi \inst{14}, D. Y. Tsvetkov \inst{15}, M. Turatto \inst{4}, S. Wilke \inst{11}
  }

\institute{
$^{1}$  Astrophysics Research Centre, School of Mathematics and Physics, Queen's University Belfast, BT7 1NN, United Kingdom
\newline
$^{2}$  Astrophysics Group, Blackett Laboratory, Imperial College London, Prince Consort Road, London SW7 1AZ, United Kingdom
\newline
$^{3}$  Max-Planck-Institut f\"{u}r Astrophysik, Karl-Schwarzschild-Str. 1, 85741 Garching bei M\"{u}nchen, Germany 
\newline
$^{4}$  INAF - Osservatorio Astronomico, vicolo dell'Osservatorio 5, 35122, Padova, Italy
\newline
$^{5}$  Pulkovo Central Astronomical Observatory, Pulkovskoe shosse 65, 196140, St. Petersberg, Russia
\newline
$^{6}$  Astronomical Institute of St Petersburg State University, Universitetskij Prospect 28, Petrodvorets, 198504 St Petersburg, Russia
\newline
$^{7}$  INAF Osservatorio Astronomico di Collurania, via M. Maggini, 64100 Teramo, Italy
\newline
$^{8}$  Spitzer Science Center, California Institute of Technology, 1200 E.California Blvd., Pasadena, CA 91125, USA.
\newline
$^{9}$  
Institut for Fysik og Astronomi, $\AA$rhus Universitet, Ny Munkegade, Bygn. 1520, 8000 $\AA$rhus C 
\newline
$^{10}$  Dark Cosmology Centre, Niels Bohr Institute, University of Copenhagen, Juliane Maries Vej 30, 2100 
Copenhagen, Denmark 
\newline
$^{11}$  Universit\"{a}ts-Sternwarte M\"{u}nchen, Scheinenstr. 1 81679 M\"{u}nchen, Germany
\newline
$^{12}$  Max-Planck-Institut f\"{u}r Extraterrestrische Physik, Giessenbachstr, 85748 Garching bei M\"{u}nchen, Germany
\newline
$^{13}$  Scuola Normale Superiore, Piazza Cavalieri, 7, 56126 Pisa, Italy
\newline
$^{14}$  Joint Astronomy Centre, 660 N. A'ohoku Place, Hilo, HI 96720
\newline
$^{15}$
Sternberg Astronomical Institute, University Ave. 13, 119992 Moscow, Russia
}

\offprints{Deborah J. Hunter, \email{dhunter$07$@qub.ac.uk}}

\date{Received / Accepted }

\titlerunning{SN 2007gr}
\authorrunning{Hunter {\em et al.}}

\abstract
{
We present photometric and spectroscopic observations at optical and near-infrared wavelengths of the nearby type Ic supernova~2007gr. These represent the most extensive data-set to date of any supernova of this sub-type, with frequent coverage from shortly after discovery to more than one year post-explosion.  We deduce a rise time to $B$-band maximum of $11.5 \pm 2.7\,$d.  We find a peak $B$-band magnitude of M$_{B}=-16.8$, and light curves which are remarkably similar to the so-called 'hypernova'  \object{SN~2002ap}. In contrast, the spectra of SNe~2007gr and 2002ap show marked differences, not least in their respective expansion velocities. We attribute these differences primarily to the density profiles of their progenitor stars at the time of explosion i.e. a more compact star for SN~2007gr compared to SN~2002ap. From the quasi-bolometric light curve of SN~2007gr, we estimate that 0.076 $\pm$ 0.010\,$M_\odot$ of $^{56}$Ni was produced in the explosion. Our near-infrared (IR) spectra clearly show the onset and disappearance of the first overtone of carbon monoxide (CO) between $\sim$\,\,70 to \,175\,d
relative to $B$-band maximum. 
The detection of the CO molecule implies that ionised He was not microscopically mixed within the carbon/oxygen layers. From the optical spectra, near-IR light curves, and colour evolution, we find no evidence for dust condensation in the ejecta out to about +\,400\,d.
Given the combination of unprecedented temporal coverage, and high signal-to-noise data, we suggest that SN~2007gr could be used as a template object for supernovae of this sub-class.

	}

\maketitle

\section{Introduction}
\label{sec:intro}

One of the main objectives of supernovae (SNe) research is to understand the relation between the physics of the explosion and the nature of the progenitor star and its immediate environment.

Type I SNe, i.e. those lacking spectroscopic signatures of hydrogen, are divided into 
the Ia, Ib, and Ic categories. Of these, type Ia SNe constitute a physically distinct class and are believed to arise from the thermonuclear explosion of an accreting carbon-oxygen white dwarf.  Instead, both the type Ib (He-rich) and the type Ic (He-poor) SNe are believed to originate from the core-collapse of stars more massive than $\sim$\,8$M_\odot$. SNe of type Ib and type Ic are quite heterogeneous in their spectral properties and energetics, which range from the relatively low kinetic energy ($\sim$10$^{51}$ erg) events like \object{SN~1994I} \citep{Nomoto94}, to the high kinetic energy ($\sim$10$^{52-53}$ erg) broad-lined events sometimes dubbed `hypernovae' e.g. the broad-lined type Ic~\object{SNe~1998bw} \citep{Galama98,Iwamoto98} and \object{2003dh} \citep{Hjorth03,Mazzali03}.  These latter SNe have been associated with long-duration gamma-ray bursts (GRBs) \citep[see][for a recent review]{Woosley06}.

At present, there is no consensus regarding the differences between the progenitors that 
produce type Ib and Ic SNe. However, it is generally accepted that Wolf-Rayet stars are 
the most promising candidates as these have shed their outer layer of H, as well
as varying amounts of their He layer \citep{Wheeler85}. The H and He layers are primarily shed through strong stellar winds, resulting in WC and WO stars (i.e. carbon and oxygen dominated, respectively). The same mass-loss mechanism may likewise only remove the H layers so that the potential progenitors of type Ib SNe i.e. WN stars (nitrogen dominating over carbon) exhibit the products of core-H burning.  Alternatively, the outermost layers may be stripped off by a companion star \citep{Podsiadlowski92} with the degree of envelope-stripping depending on the configuration of the binary system \citep{Pols97}.

The distinction between type Ib and type Ic SNe is historical, and challenged by the idea that He may be present in both types. The detection of spectral features due to He in type Ib SNe might simply imply that the element is high in abundance and/or a sufficient quantity of radioactive nickel is mixed into the He layer, providing an excitation source \citep{Wheeler87,Shigeyama90,Hachisu91}. The converse may be true for type Ic SNe explaining the non-detection of optical He~I lines in type Ic spectra \citep{Nomoto90,Hachisu91}.  Most important, no type Ic SN has yet displayed the isolated near-IR He line at 2.058\,$\mu$m to a strength comparable to those detected in type Ib SNe.

Consequently, it is possible that a continuous degree of He abundances and/or excitation exists from type Ib to Ic SNe, making the distribution smooth rather than bimodal. This seems consistent with the identification of intermediate cases such as \object{SN~1999ex} which was characterised by weak optical He I lines, and strong He I $\lambda$$\lambda$ 10830, 20581 lines in the near-IR \citep{Hamuy02}.  Additionally, the transitional nature of SNe such as \object{SN~2005bf} \citep{Folatelli06} and \object{SN~2008D} \citep[][]{Soderberg08,Mazzali08,Malesani09}, both of which underwent a metamorphosis from a type Ic at early times to a type Ib at later epochs, may also provide an insight into the controversy of He.  The current challenge lies in linking the observed variations of type Ib/c SNe to the physical properties of their progenitor systems.

This paper presents the results of an intensive observational follow-up campaign of the type Ic SN~2007gr from shortly after explosion to more than a year later.  The layout of this paper is as follows: in section \ref{sec:SN2007gr} we provide some basic information on SN~2007gr and its host galaxy, with the data acquisition and reduction procedures described in section 3. This is followed in section 4 by a description of the optical and near-IR photometric behaviour, including the colour evolution and quasi-bolometric light curve. Section 5 is devoted to the spectroscopic evolution; we end with a summary in section 6.

\section{SN~2007gr}\label{sec:SN2007gr}

SN~2007gr was discovered during the course of the Lick Observatory Supernova Search 
(LOSS) on 2007 August 15.51 UT \citep{Madison07}. The SN exploded about 24$\farcs$8 W and 15$\farcs$8 N of the nucleus of the nearby spiral galaxy NGC~1058.  \citet{Chornock07} classified SN~2007gr to be of type Ib/c from a spectrum taken soon after explosion.  The classification was later refined to type Ic as the presence of He could not be confirmed in subsequent spectra \citep{Valenti08a}.

Given that the object was undetected in an image taken on 2007 August 10.44 \citep[unfiltered mag $<$ 18.9;][]{Madison07}, the explosion date can be confined to less than 5 days before discovery. Here, we adopt 2007 August 13 (JD 2,454,325.5 $\pm$ 2.5) as the explosion epoch, and 2007 August 25 (JD 2,454,337.0 $\pm$ 1.0) as the time of $B$ band maximum, used hereafter as the reference epoch, t\,=\,0 days (see section \ref{sec:Optical_LC}). Although no source was detected at the location of the SN in HST pre-explosion images \citep{Crockett08}, the SN was suggested to be a cluster member based on its close location to a bright source.  The pre-explosion HST and ground based images marginally favour a young cluster with a turn-off mass of~28 $\pm$~4$~M_\odot$. 

The host galaxy of SN~2007gr, NGC~1058, belongs to a group of nearby galaxies of which NGC~925 is also a member.  The distance to NGC~925 has been derived by \citet{Silbermann96} as 9.29 $\pm$ 0.69 Mpc ($\mu$\,=\,29.84 $\pm$ 0.16) using Cepheid variables; we adopt this value throughout the paper for the distance to SN~2007gr. We note that an alternative distance estimate based on the Tully-Fisher relation has also been reported \citep[10.6 $\pm$ 1.3 Mpc;][]{Terry02}.  The galaxy of SN~2007gr also hosted two prior supernovae, \object{SN~1969L} \citep[$A$$_{V}$\,=\,0.17;][]{Burstein84} and \object{SN~1961V} \citep[$A$$_{V}$\,$\sim$ 2.2-4.1;][]{Filippenko95a} with the latter possibly being the outburst of a luminous blue variable \citep{Goodrich89}.  

We adopt a galactic extinction of $E(B-V)$\,=\,0.062 \citep{Schlegel98}, and a host galaxy 
extinction of $E(B-V)$\,=\,0.03 which was estimated using the equivalent widths of Na I D absorption lines in a medium-resolution spectrum \citep{Valenti08a}.  We assume a total uncertainty of 20\% for our estimate of the total extinction $E(B-V)$$_{tot}$\,=\,0.092 $\pm$ 0.018 mag.

\section{Data acquisition and reduction}\label{sec:reduction}

The proximity of SN~2007gr lent itself to an extensive observational campaign starting
6 days before maximum light and extending through to +404 days. The data acquisition
and reduction procedures are detailed below. 

\subsection{Photometry}\label{sec:photometry}

$UBVRI$ photometry of SN~2007gr was obtained from 2007 Aug 18 to 2008 Sep 30, spanning approximately $-\,6$ to $+\,404$ days with respect to $B$-band maximum.

The basic reduction of the data i.e., trimming, bias and overscan correction, and flat-fielding
was performed using standard procedures within the IRAF environment. Images taken on the same night with the same filter were combined to improve the signal-to-noise ratio.

\begin{figure}[t]
\begin{center}
\includegraphics[width=0.49\textwidth]{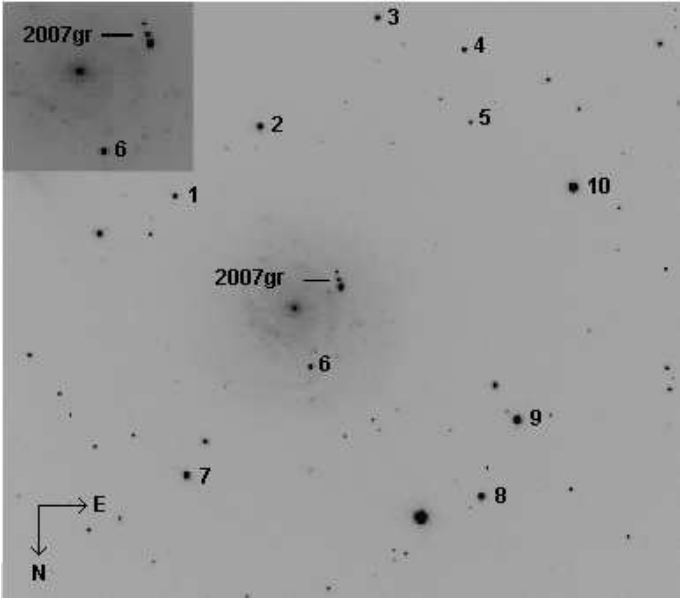}
\caption{Section of $V$-band image taken on 2008 Jan 13 ($\sim+\,142.4$\,d relative to $B$ maximum) with the Nordic Optical Telescope (NOT). 
SN~2007gr and the local sequence of stars are marked. The field of view is $6'.3$ $\times 6'.3$.  
The inset shows the two bright stars that flank SN~2007gr, both of which are approximately 5$''$ from the SN. }
\label{fig:Field}
\end{center}
\end{figure}

The instrumental magnitudes of SN~2007gr were obtained via the PSF fitting technique as implemented in the SNOOPY\footnote{SNOOPY was originally devised by \citet{Patat96} and implemented in IRAF by Cappellaro. The package is based on DAOPHOT and has been optimized for SNe.} software package.  This technique was preferred over aperture photometry because of the complex location of SN~2007gr, close to two bright sources (see inset Fig. \ref{fig:Field}). 

Only for the observations taken on 2008 Feb 10, did we subtract a template image of the host galaxy before performing our measurements on the SN. This was necessary given the  poor seeing conditions ($>1\farcs8$). For the template, we used archival images obtained on 2005 Jan. 13 with the Isaac Newton Telescope. 

The galaxy subtraction was performed using a purpose-written script based on the ISIS package \citep{Alard00,Alard98} and implemented into the SNOOPY package.  Template subtraction was not applied to all the data as the only template images available, with similar filters to the Landolt system (Johnson-\,Cousins), were those from the Isaac Newton Telescope described above. The SN photometry was performed relative to a local sequence of 10 stars in the field of NGC~1058, calibrated during several photometric nights by comparison with Landolt standard stars (Fig. \ref{fig:Field}).  The $UBVRI$ magnitudes of the local standards are reported in Table \ref{local_stars_mags}.

In spite of the many different instrumental configurations used, the colour-corrected $UBVR$ 
photometry of SN~2007gr appears to be coherent. In contrast, the raw $I$-band light curve showed a significant scatter (up to 0.4 mag) due to the differences in the $I$ filter transmission curves between the various instruments.  We therefore applied an S-correction to the $I$ band data \citep{Pignata08} in order to standardise the photometry. This technique has previously been applied to a number of type Ia SNe \citep[e.g.][]{Stritzinger02,Pignata04,Pignata08}. Application of the S-correction to the $I$-~band data substantially reduced the scatter in the light curve.  The S-corrected light curves of SN~2007gr are displayed in Fig. \ref{fig:Optical_LC_07gr} and the magnitudes are reported in Table A.2.

Near-IR photometry in the $JHK$ bands was obtained on 25 nights over a period of more than one year (Table \ref{NIR_photometry}). The near-IR magnitudes were determined using the same PSF-fitting technique as described for the optical photometry. The SN magnitudes were calibrated with respect to field stars in the 2MASS~\footnote{http://www.ipac.caltech.edu/2mass/index.html} database (see section \ref{sec:NIR_LC} for light curves).  The derived $JHK$ magnitudes and their uncertainties are listed in Table~\ref{NIR_photometry}.  The first-order colour corrections were omitted when observations in one band only were obtained.

\subsection{Spectroscopy}\label{sec:spectroscopy}

Optical spectra were obtained at 29 epochs, spanning $-\,7.3$ to $+\,375.5$ days with respect to $B$ maximum. Table \ref{log_optical_spectra} lists the dates, instrument configurations, as well as the resulting spectral resolution and wavelength coverage.  

The 2D spectral images were processed in a standard fashion using IRAF, including bias subtraction and flat-field corrections. The spectra were wavelength calibrated using arc lamps; the calibration was checked against night-sky lines and if required small shifts ($\lesssim$ few $\AA$) were applied to the SN spectra. The spectra were flux calibrated using spectrophotometric standard-star spectra taken with an identical instrument configuration.  All observations were obtained with a slit angle chosen to minimise the contribution of the two stars flanking SN~2007gr (see inset Fig. \ref{fig:Field}). The slit was therefore not necessarily aligned along the parallactic angle. However, the spectra were corrected for light losses by 
making use of the extensive multi-band photometry available.

Our last optical spectrum was taken on 2008 Sep. 3, approximately 387 days after explosion, with the Gemini Multi-Object Spectrograph on the Gemini North Telescope (GMOS-N). The reduction procedure for this spectrum differed slightly compared to the other spectra as we processed the data with the GMOS pipeline implemented within IRAF.  The spectrum revealed significant contamination from nearby bright stars at bluer wavelengths.  The spectra of these nearby sources were extracted with the SN and their flux contributions were subsequently subtracted from the SN spectrum using standard tasks within IRAF.  The optical spectra of SN~2007gr are discussed in section \ref{sec:Optical_spectra}.

SN~2007gr was first observed in the near-IR on 2007 Sep 9 UT (+\,15.5\,d), and subsequently 
ten more times. The bulk of our near-infrared spectroscopic data were acquired using the near-IR imager and spectrometer (NIRI) mounted at the Cassegrain focus of the GEMINI-North Telescope.  Three epochs were obtained using the Near Infrared Camera Spectrometer (NICS) at the TNG. Details of these observations, including instrument configurations, are given in Table \ref{log_NIR_spectra}.

The NIRI data were processed using the GEMINI/NIRI pipeline implemented within the IRAF environment, while the TNG data were processed using standard routines in IRAF. All of our near-IR spectra were acquired by nodding the point-source targets along the spectrograph slit in the standard ABBA pattern, facilitating the removal of background emission.  For the TNG data, the sky subtraction was performed by subtracting the $A$ frames from the $B$ frames (and vice-versa), and one-dimensional spectra of SN~2007gr were then extracted from the sky-subtracted frames.  In the case of the Gemini data, the AB frames were pairwise subtracted and divided by the normalised flat field.  A sky frame was then derived for each individual exposure by averaging the nearest 6-10 dithered frames taken within 2 minutes.  At later epochs, the time period was set in such a way so that one sky frame was selected per object image obtained on 2007 Dec 20 and 2008 Jan 09.  

The near-IR spectra were all wavelength-calibrated using arc lamps. We used spectra of standard stars, taken at similar air masses and on the same nights as the SN spectra, to remove telluric absorptions and to obtain a relative flux calibration. The final flux calibration of the spectra was achieved by applying scaling factors derived from the near-IR photometry (Table \ref{NIR_photometry}).  Our series of near-IR spectra are shown in section \ref{sec:nir_analysis}.

The spectra presented are in the galaxy rest-frame assuming a redshift of the galaxy NGC~1058 i.e. $z$\,= \,0.001728 \footnote{The redshift of NGC~1058 is taken from NED (Nasa/Ipac Extragalactic Database).}.  From our latest optical spectrum on 2008 Sep 03 (+\,375.5\,d),
we measure a redshift from the H$\alpha$ line due to NGC~1058 that is consistent with the value reported above. The spectra are corrected for extinction using the standard extinction law of \citet{Cardelli89} and $E(B-V)_{tot}$\,=\,0.092 mag (see section \ref{sec:Optical_LC}).

\section{Photometry}
\subsection{Optical Light Curves} \label{sec:Optical_LC}

\begin{figure}[t]
\centering 
\includegraphics[width=0.5\textwidth,angle=0.0,trim=5mm 0mm 0mm -1mm,clip]{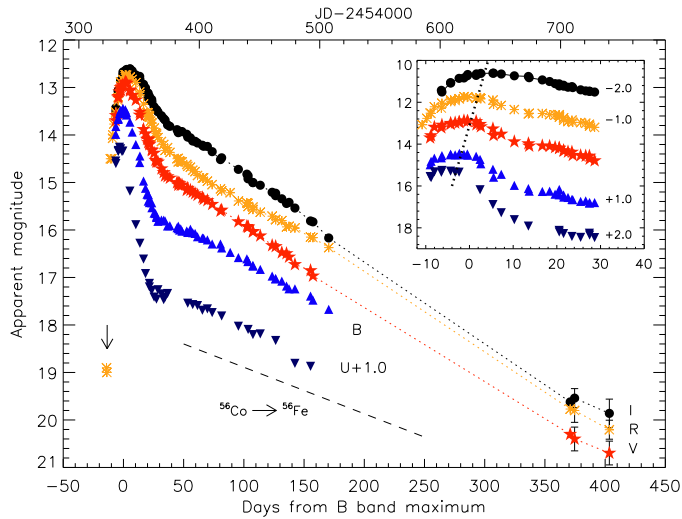}
\caption{Complete $UBVRI$ light curves of SN~2007gr.  $I$-band data obtained from the Liverpool and Wendelstein Telescopes have been S-corrected (\S \ref{sec:photometry}). All errors at phases $<$ 200\,d are smaller than the symbol sizes.
The dashed line shows the rate of decay of $^{56}$Co$\rightarrow$$^{56}$Fe. 
The arrow indicates the pre-explosion $R$-band detection limit at the location of SN~2007gr.  
The inset shows the light curve evolution at early times with the dotted line approximately linking the peak magnitudes in all bands.}
\label{fig:Optical_LC_07gr}
\end{figure}

The complete optical light curves of SN~2007gr, extending from $-6$ to $404$\,d after $B$ maximum, are displayed in Figure \ref{fig:Optical_LC_07gr}.  The dense photometric coverage of SN~2007gr allows us to estimate the epoch of maxima and the peak apparent magnitudes in all bands. This was done by fitting each light curve around maximum with a low-order polynomial function. Table \ref{07gr_parameters} lists the peak apparent and absolute magnitudes, as well as the epochs of maxima in all bands with respect to both $B$-band maximum and the date of explosion.  All photometry has been corrected for extinction using the relations of \citet{Cardelli89} and a colour excess $E(B-V)$$_{tot}$\,=\,0.092 mag (see Table \ref{07gr_parameters}).

$B$-\,band maximum occurred on 2007 August 25 at 13.47 $\pm$ 0.02 mag, with the $V$, $R$, and $I$ bands peaking respectively about 1.5, 4.0, and 4.0\,d later.  With a distance modulus $\mu$\,=\,29.84, we estimate the absolute $B$-band magnitude of SN~2007gr at maximum to be $-16.75$.  This is the same as for SN~2002ap ($M$$_{B} =-16.76$), but significantly fainter than the GRB associated SN~1998bw ($M$$_{B}=-18.44$) and SN~2003jd ($M$$_{B}=-19.30$).~\footnote{References are given in Table \ref{tab:Ic_parameters}.}  Since the peak brightness depends primarily on the mass of $^{56}$Ni synthesized in the explosion, we may immediately surmise that SNe~2002ap and 2007gr appear to have comparable quantities of $^{56}$Ni. We will discuss this further in section \ref{sec:Bolometric}.

\addtocounter{table}{0}
\begin{table*}[t]
\begin{center}
\caption{The photometric parameters of SN~2007gr.}
\begin{footnotesize}
\scriptsize
\begin{tabular}{lllllllll}
\hline\hline
&$U$ 			&$B$ 			&$V$ 			&$R$ 			&$I$				&$J$				&$H$			&$K$		\\
\hline	
		          			&                			&                			&                			&				&    			  	&				&				&			 \\
Apparent mag.  				&13.29 $\pm$ 0.02	&13.47 $\pm$ 0.02	&12.91 $\pm$ 0.01	&12.76 $\pm$ 0.02	&12.61 $\pm$ 0.02	&12.40 $\pm$ 0.18	&12.06 $\pm$ 0.10	&11.66 $\pm$ 0.10		\\
at maximum          			&                			&                			&                			&				&    			  	&				&				&			 \\
	&                			&                			&                			&				&    			  	&				&				&			 \\
Absolute mag. 				&-16.99  $\pm$ 0.19	&-16.75 $\pm$ 0.19	&-17.22 $\pm$ 0.18	&-17.29 $\pm$ 0.18	&-17.36 $\pm$ 0.17	&-17.52 $\pm$ 0.20	&-17.83 $\pm$ 0.20	&-18.21 $\pm$ 0.20		\\
at maximum  \hspace{0.5mm}\inst{a}         		&                			&                			&                			&				&    			  	&				&				&		 	\\
&       &                			&                			&				&    			  	&				&				&			 \\
Extinction (mag)  			&0.44  $\pm$ 0.09	&0.38 $\pm$ 0.08	&0.29 $\pm$ 0.06	&0.21 $\pm$ 0.04	&0.13 $\pm$ 0.03	&0.08 $\pm$ 0.02	&0.05 $\pm$ 0.01	&0.03 $\pm$ 0.01		\\
&                			&                			&                			&				&    			  	&				&				&		 	\\
JD of   			&54334.8 $\pm$ 0.8	&54337.0	$\pm$ 1.0	&54338.5 $\pm$ 1.1	&54341.0 $\pm$ 1.7	&54341.0 $\pm$ 1.5	&54341.4 $\pm$ 2.3	&54342.4$\pm$ 4.3	&54342.3 $\pm$ 3.5		\\
maximum \hspace{0.5mm}\inst{b}
	&     	&	    			&	 			&				&       			&				&		&		\\
	&       &                			&                			&				&    			  	&				&		&		 \\
Epoch of    	&-2.2 $\pm$ 1.3	&0 $\pm$ 1.0 		&1.5 $\pm$ 1.5	  	&4.0 $\pm$ 2.0		&4.0 $\pm$ 1.8		&4.4 $\pm$ 2.5		&5.4 $\pm$ 4.4		&5.3 $\pm$ 3.6 		\\
maximum (days)  \hspace{0.5mm}\inst{c}  		&     				&     				&     				&     				&      				&				&		 &		\\
		          			&                			&                			&                			&				&    			  	&				&		&		 \\
Rising        &9.3 $\pm$ 2.8		&11.5 $\pm$ 2.7 	&13.0 $\pm$ 2.9	&15.5 $\pm$ 3.2	&15.5 $\pm$ 3.1	&15.9 $\pm$ 3.0	&16.9 $\pm$ 5.1	&16.8 $\pm$ 4.4		\\
times (days)  \hspace{0.5mm}\inst{d}  				&     				&     				&     				&     				&      				&			&		 &		\\
		          			&                			&                			&                			&				&    			  	&			&		&		 \\
\hline
		          			&                			&                			&                			&				&    			  	&			&		&		 \\
$\bf{Decline}$ $\bf{rates}$  \hspace{0.5mm}\inst{e}     	&     				&	    			&	 			&				&      				&			&		&		\\
(30-100d)  \hspace{0.5mm}\inst{c} 					&0.0095 $\pm$ 0.0007	&0.0111 $\pm$ 0.0006	&0.0170 $\pm$ 0.0005	&0.0194 $\pm$ 0.0007	&0.0164 $\pm$ 0.0004	&0.0304 $\pm$ 0.0006		&0.0268 $\pm$ 0.0008	&0.0342 $\pm$ 0.0008		\\
(100-170d)  \hspace{0.5mm}\inst{c} 				&0.0169 $\pm$ 0.0011	&0.0153 $\pm$ 0.0003	&0.0189 $\pm$ 0.0003	&0.0128 $\pm$ 0.0004	&0.0167 $\pm$ 0.0005	&$\ldots$		&$\ldots$	&$\ldots$		\\
(30-170d)  \hspace{0.5mm}\inst{c} 					&0.0114 $\pm$ 0.0006	&0.0130 $\pm$ 0.0004	&0.0171 $\pm$ 0.0003	&0.0159 $\pm$ 0.0005	&0.0169 $\pm$ 0.0002 	&$\ldots$		&$\ldots$	&$\ldots$		\\
		          			&                			&                			&                			&				&    			  	&			&		&		 \\
\hline
\end{tabular}
\\[1.6ex]
\hspace{20 mm} \inst{a}  A distance modulus $\mu$\,=\,29.84 $\pm$ 0.16 mag and a colour excess E(B-V)\,=\,0.092 mag were adopted.
\newline
\hspace{0 mm} \inst{b}  The uncertainty in the epoch of maximum is defined as the time period in which the magnitude is not more than 0.01 mag fainter than the peak magnitude \citep{Yoshii03}.
\newline
\hspace{0 mm} \inst{c}  Epoch with respect to the estimated $B$-band maximum JD 2,454,337.0 $\pm$ 1.0.
\hspace{2 mm} \inst{d}  Epoch with respect to the estimated explosion date JD 2,454,325.5 $\pm$ 2.5.
\newline
\inst{e}  Decline rates (mag d$^{-1}$) were estimated using linear least-square fitting.
\end{footnotesize}
\label{07gr_parameters}
\end{center} 
\end{table*}

The rise time of SN~2007gr, hence likely the time interval between explosion and maximum is constrained by the $R$-band non-detection on 2007 August 10 \citep{Madison07}. This restricts the rise time to $B$ maximum to 11.5\,$\pm$\,2.7\,d (i.e. 8.8\,-\,14.2\,d), indicating that SN~2007gr has an intermediate rise time compared to other type Ic SNe. It is typical to those of SN~2002ap \citep[8\,d;][]{Foley03}, \object{SN~1998bw} \citep[$\sim$15\,d;][]{Mazzali01}, and SN~1994I which reached $B$ maximum $\sim$12 days after explosion \citep{Iwamoto94}. 

We estimated the decline rates for SN~2007gr by using linear fits to the light curves (see Table \ref{07gr_parameters}).  The $UBVRI$ light curves reveal a rapid decline after maximum followed by a much slower decline, which is typical of SNe Ib/c.  The decline rate in the period from 30 to 170\,d after the explosion (i.e. $\sim$ 0.01-\,0.017 mag d$^{-1}$) is similar in all optical bands, but typically a bit faster in the redder bands.  The decline in the earlier part (30-\,100\,d) is generally less steep than at later epochs (100-\,170\,d) with the exception of the $R$ band.   The $R$ light curve initially declines faster than other optical bands and is followed by a phase of slower decline from 100 to 170\,d. This is probably due to the development of forbidden [Ca II] + [O II] $\sim$~7300 $\AA$ and [O I] $\lambda$$\lambda$ 6300, 6364 arising at $\sim$\,80\,d, as in the case of typical SNe Ib/c.  At epochs later than 60 days, the light curve slopes of SN~2007gr in all optical bands are found to be steeper than those expected when the energy source is $^{56}$Co$\rightarrow$$^{56}$Fe decay and the trapping of $\gamma$-rays is complete [0.98 mag (100 d)$^{-1}$] (see Fig.\ref{fig:Optical_LC_07gr}). This is probably indicative of significant $\gamma$-ray escape.

\begin{figure*}[t]
\centering
\includegraphics[width=1.00\textwidth,angle=0.0,trim=0mm 0mm 0mm 0mm,clip]{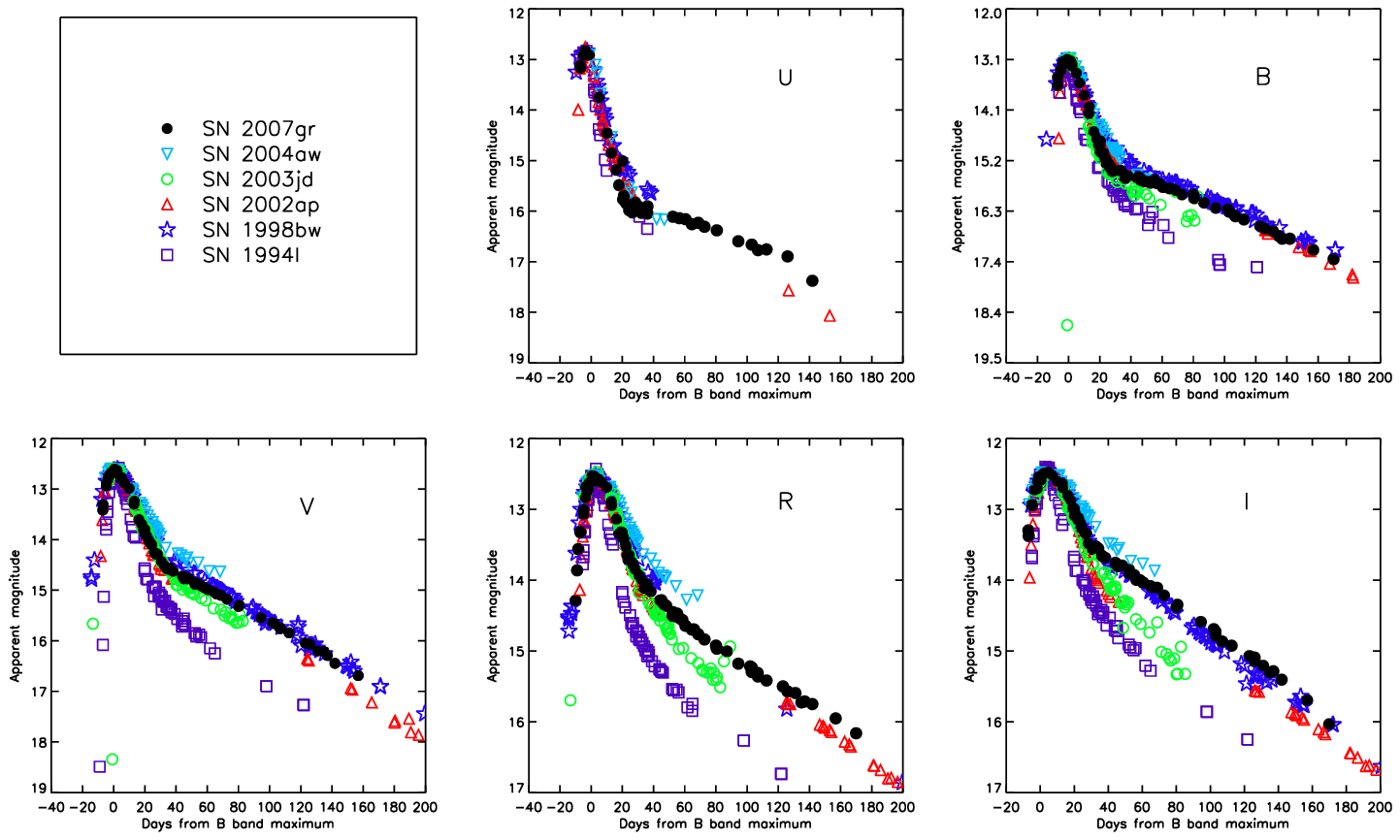}
\caption{Optical light curves of SN~2007gr compared to those of the Type Ic \object{SNe~2004aw} \citep{Taubenberger06}, \object{2003jd} \citep{Valenti08b}, 2002ap \citep{Pandey03,Foley03,Tomita06}, 1998bw \citep{Galama98,McKenzie99} and 1994I \citep{Yokoo94,Gyoon95,Lee95,Tsvetkov95,Richmond96}.  
All the light curves presented have been shifted in time and magnitude to match SN~2007gr at maximum. }
\label{fig:Optical_LC_Ic}
\end{figure*}

In Figure \ref{fig:Optical_LC_Ic} we compare the $UBVRI$ light curves of SN~2007gr with those of other well observed stripped core-collapse SNe Ic.  These SNe represent an exhaustive sample of type Ic SNe with well-sampled light curves; luckily, these SNe also span a large range of properties, thereby probing different physical conditions.

The light curves of the SNe have been scaled in time and magnitude to match the peaks of SN~2007gr in all filters.  Figure \ref{fig:Optical_LC_Ic} illustrates the diversity of SNe Ic in terms of their light curve morphology and decline rates.  Although \cite{Clocchiatti97} suggest that the decline rates appear to come in $\it{slow}$ and $\it{fast}$ varieties, new events seem to support the idea that there is a continuous range between these two extremes.
Fig. \ref{fig:Optical_LC_Ic} shows that after maximum brightness, SN~1994I is the fastest decliner in all filters followed by SN~2003jd, while SN~2004aw is generally the slowest. In the nebular phase, the decline rates ($\gamma$$_{LC}$) of the light curves are determined by the energy deposition of the $\gamma$-rays produced in the $^{56}$Ni$\rightarrow$$^{56}$Co$\rightarrow$$^{56}$Fe radioactive decay chain.  Thus the emitted luminosity includes the $\gamma$-ray contributions which result from the energy deposition in the $^{56}$Co decay and from the annihilation of electron-positron pairs, in which the kinetic energy of the positrons is an important factor.  Assuming homologous expansion and the incomplete trapping  of $\gamma$-rays and positrons, the behaviour of the light curve depends on the mass ejected ($M$$_{ej}$) and the kinetic energy ($E$$_{kin}$) according to the following relation (see \citet{Valenti08b} and references therein for a simple derivation),
\begin{equation}
\hspace{3.25 cm} \gamma_{LC} \propto M_{ej} / {E_{kin}}^{1/2}
\end{equation}
Hence, fast decliners such as SN~1994I ($\gamma$$_{LC}$\,$\sim$\,0.88) are typical of SNe with lower ratios of ejecta mass to kinetic energy than the slow decliners of SN~2004aw ($\gamma$$_{LC}$\,$\sim$\,2.3) and SN~1998bw ($\gamma$$_{LC}$\,$\sim$\,1.8) (see Table \ref{tab:Ic_parameters}).  Meanwhile, the $U$ band of SN~2007gr fades at a rate similar to those of SNe~2004aw and 2002ap from $B$ maximum to $\sim$50\,d later, while the later evolution is slower than in SN~2002ap. For the $B$, $V$ and $R$ bands, the decline is very similar to SNe~2002ap and 1998bw, and faster than SN~2004aw.

\addtocounter{table}{0}
\begin{table*}[t]
\begin{center}
\caption{The parameters of type Ic SNe}
\label{tab:Ic_parameters}
\begin{footnotesize}
\scriptsize
\begin{tabular}{llllllllllll}
\hline\hline
SN (Ic)  	& $^{56}$Ni mass 			& Ejecta mass 	& E$_{kin}$ 		& m$_{B,max}$		& M$_{Bmax}$$^{a}$	 	&Rise			&E(B-V)$^{b}$	&$\mu$	&Redshift$^{c}$	&Ref	\\
		&($M_{\sun}$)			&($M_{\sun}$)	&(10$^{51}$ ergs)	&				&				&times (d)			&			&		&				&	\\
\hline
2007gr	&  0.076 $\pm$  0.020		&  2.0-3.5		&  1-4			&13.47		&-16.75				&11.5 $\pm$ 2.5	&0.092		&29.84	&0.001728		&1, 2	\\
2004aw   	&  0.30 $\pm$  0.05		&  3.5-8.0		&  3.5-9.0			&18.06		&-17.63				&$\ldots$		&0.370		&34.17	&0.015914		&3	\\
2003jd     	&  0.36 $\pm$ 0.04		&  2.5-3.5	 	&  5-10			&15.75		&-19.30				&$\sim$13		&0.144		&34.46	&0.018860		&4	\\
2002ap	&  0.09, 0.07			&  2.5-5.0	 	&  4-10			&13.11		&-16.76				&$\sim$8		&0.090		&29.50	&0.002192		&5, 6, 7, 8, 9	\\
1998bw	&  0.7, 0.4, 0.5			&  10.9		&  20-50			&14.09		&-18.44				&$\sim$15		&0.060		&32.28	&0.008670		&10, 11, 12, 13  \\
1997ef	&  0.13			&  9.6			&  17.5			&17.45       	&-16.18				&$\ldots$		&0.000		& 33.63	&0.011688		&14	\\
1994I		&  0.07				&  0.88		&  1				&13.77		&-17.06				&$\sim$12		&0.300		&29.60	&0.001544		&15, 16, 17	\\
\hline
\end{tabular}
\\[1.6ex]
References: \hspace{2 mm}\inst{1}  \cite{Valenti08a}		\hspace{2 mm}\inst{2}  This paper		\hspace{2 mm}\inst{3}  \cite{Taubenberger06}		\hspace{2 mm}\inst{4}  \cite{Valenti08b}
		    \hspace{2 mm}\inst{5} \cite{Yoshii03}		\hspace{2 mm}\inst{6} \cite{Foley03}		\hspace{2 mm}\inst{7} \cite{Gal-Yam02}		\hspace{2 mm}\inst{8} \cite{Tomita06}	\hspace{2 mm}\inst{9} \cite{Mazzali02}			\hspace{2 mm}\inst{10} \cite{Iwamoto98}		\hspace{2 mm}\inst{11} \cite{Galama98}			\hspace{2 mm}\inst{12} \cite{Nakamura01}	\hspace{2 mm}\inst{13} \cite{Patat01}		\hspace{2 mm}\inst{14} \cite{Mazzali00}	\hspace{2 mm}\inst{15} \cite{Nomoto94}		\hspace{2 mm}\inst{16} \cite{Richmond96}	\hspace{2 mm}\inst{17} \cite{Sauer06}  
\hspace{2 mm} $^{a}$ The peak absolute magnitudes were estimated using 
the apparent magnitudes at maxima. 
\hspace{2 mm} $^{b}$ Total extinction   \hspace{2 mm} $^{c}$ Redshift of the host galaxy (taken from the Nasa/IPAC Extragalactic Database \textit{NED}).
\end{footnotesize}
\end{center} 
\end{table*}

\subsection{Near-IR Light Curves}\label{sec:NIR_LC}

Our near-IR photometry for SN~2007gr is reported in Table \ref{NIR_photometry}.  Early observations in the $J$-band enable us to constrain the peak magnitude and epoch of $J$ maximum.  This is likely to have occurred on 2007 August 29, $\sim$16 days after the explosion date and 4.4 days after $B$ maximum.

\begin{figure}[t]
\centering
\includegraphics[width=0.5\textwidth, trim=4mm 1mm 0mm -1mm, clip]{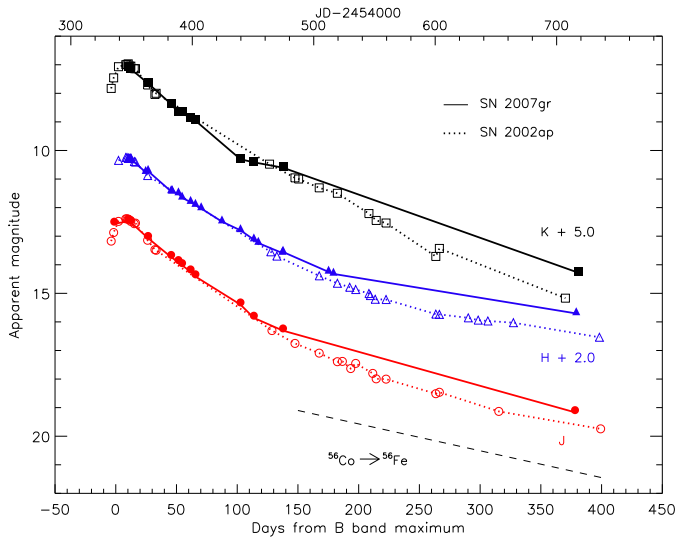}
\caption{The near-IR light curves of SNe~2007gr and 2002ap \citep{Tomita06}.  
The filled symbols joined by the full lines represent the data for SN~2007gr.
The data for SN~2002ap (open symbols, dotted lines) have been shifted in magnitude by 0.7, 0.7 and 0.5 mag to match SN~2007gr at $\sim$\,20\,d in the JHK bands, respectively.   }
\label{fig:NIR_LC}
\end{figure}

In Figure \ref{fig:NIR_LC} we compare the $JHK$ lightcurves of SN~2007gr with those of SN~2002ap \citep{Tomita06}, which has the best coverage of all SNe Ic at near-IR wavelengths. In the near-IR, the light curves of SN~2007gr are very similar to those of SN~2002ap in terms of light curve morphology.  Approximate peak magnitudes have been derived using the additional coverage of SN~2002ap as a template, particularly for the $H$ and $K$ bands (see Table \ref{tab:Ic_parameters}) where the rising branch is unconstrained. The estimated peaks of SN~2007gr suggest that the time delay between $B$ and $J$ maximum is comparable to SN~2002ap ($\sim$5 days for SN~2007gr and 7 days for SN~2002ap). The estimated decline rates for the period from +\,30 to +\,140\,d after $B$ maximum in the $J$, $H$ and $K$ bands are 0.029, 0.024 and 0.026\,mag d$^{-1}$, respectively, again similar to those of SN~2002ap.  However, the steeper decline rates of SN~2002ap at later epochs ($>$\,+\,120\,d) indicate a greater $\gamma$-ray leakage from the radioactive decay than for SN~2007gr.

\subsection{Colour Evolution of SN~2007gr}\label{sec:Colours}

\begin{figure*}[t]
\centering
\includegraphics[width=0.98\textwidth,clip=]{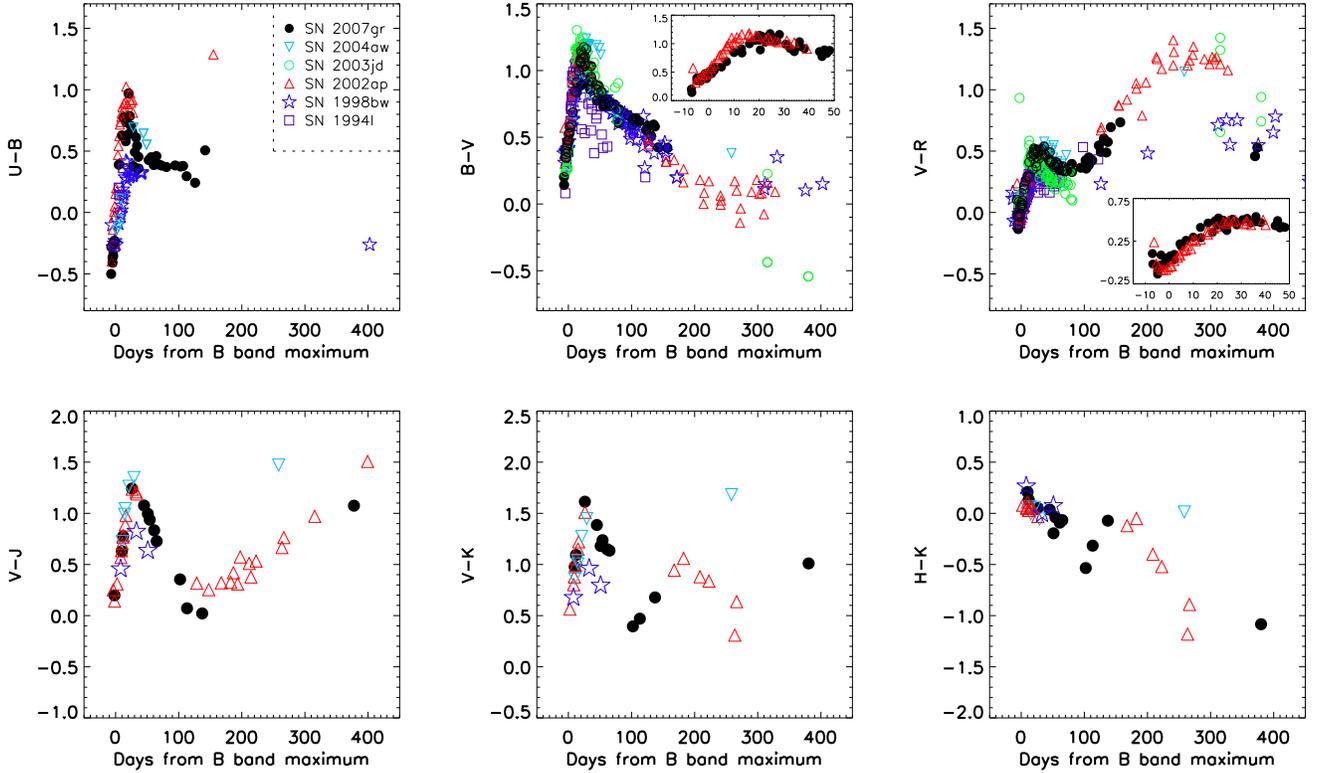}
\caption{Optical and near-infrared colour evolution of SN~2007gr compared with other SNe Ic: SN~2002ap, SN~2004aw, 
SN~1998bw, SN~1994I. Evolution of the $U-B$ (top-left), $B-V$ (top-middle), $V-R$ (top-right), $V-J$ (bottom-left), $V-K$ (bottom-middle) and $H-K$ (bottom-right) colours. The insets in two of the upper panels show a zoom-in of the early colour evolution of SN~2007gr compared to SN~2002ap.
All curves have been corrected for reddening.  
The \citet{Cardelli89} law was used to estimate the extinction in the different bands. 
See text for references.}
\label{fig:5}
\end{figure*}

The colour curves of SN~2007gr were produced using the optical and near-IR photometry detailed above.  In Figure \ref{fig:5}, we present the $U-B$, $B-V$, $V-R$, $V-J$, $V-K$ and $H-K$ colour curves of SN~2007gr and comparisons with the same sample of type Ic SNe: 2004aw, 2003jd, 2002ap, 1998bw and 1994I.  All curves have been corrected for reddening using the relations given by \cite{Cardelli89} [see Table \ref{tab:Ic_parameters} for values and references]. 

The colour evolution of all SNe presented become monotonically redder from $B$ maximum to $\sim$ +\,20\,d.  
After this time, the differences between the SNe become evident. 

The early $U-B$ colour of SN~2007gr (Fig. \ref{fig:5}, top-left panel) shows a steep rise by $\sim$\,1.5 mag.  This increasingly redder colour may be due to both the cooling of the continuum and the progressive increase of the metal line blanketing in the $U$ band. After the peak, the $U-B$ colour of SN~2007gr gradually turns bluer, followed by a plateau for $\sim$\,60\,d. Intermediate-phase data for SN~2002ap are lacking, but at t\,=\,150 days it appears to have reddened considerably by $\sim$\,0.3 mag with respect to the colour at $\sim$ +\,20\,d. SN~2007gr also  marginally reddens around this epoch.

The $B-V$ colour indices of all SNe (Fig. \ref{fig:5} , top-middle panel) in this sample rise at an approximate linear rate for the first 3 weeks before the $B$ band maximum. After the colour peak, they rapidly become bluer, except for SN~2004aw which remains the reddest at 1.2 mag ($\sim$ +\,50\,d) before showing any sign of a decline. In the case of SN~2004aw, it is difficult to ascertain whether this behaviour is intrinsic to the SN, or simply due to an underestimate of the reddening as has already been suggested by \cite{Benetti06}.  On the contrary, SN~1994I remains the bluest SN ($B-V$ $\sim$ 0.2\,-\,0.5 mag at $\sim$ +\,50\,d) throughout its evolution.  The $B-V$ colour of SN~2007gr is intermediate between these two objects, with strong similarity to SNe~1998bw and 2002ap. Uncertainties in the reddening estimates and possible contamination from nearby blue stars preclude a further discussion of the bluer colours of SN~1994I \citep{Taubenberger06}.
 
The early $V-R$ colour evolution ($<$ +\,80\,d) of SN~2007gr is comparable to the evolution of SN~2004aw with a fast rise to maximum at $V-R$ $\sim$ 0.5 mag after $\sim$ +\,20\,d. Following the peak, all SNe tend to become bluer for a short period until $\sim$ +\,70\,d. This can be explained
in part by cooling due to ejecta expansion, with the peak of the spectral energy distribution shifting towards longer wavelengths.
After this period, SN~2007gr gradually reddens, following the evolution of SN~2002ap which reaches a $V-R$ colour of $\sim$\,1.3 mag at $\sim$ +\,300\,d.  The onset of this reddening is consistent with the development of the forbidden [O I] $\lambda$$\lambda$ 6300, 6364 emission and the blend of [Ca II] $\lambda$$\lambda$ 7291, 7323 and [O II] $\lambda$$\lambda$ 7320, 7330 (see Fig. \ref{fig:Optical_spectra}).

In Figure \ref{fig:5}, the $V-J$ and $V-K $ colour curves of all SNe become monotonically redder, reaching a peak at $\sim$ +\,30\,d.  This is then followed by a rapid decline to the blue up to +\,150\,d before undergoing a further transition to a redder colour.  In the near-IR too, the colour evolution of SN~2007gr best resembles SN~2002ap.  The $V-J$ colours of SNe~2007gr and 2002ap redden to $\sim$\,1.5 mag at $\sim$ +\,400\,d.

Signatures of any dust that might have formed in the ejecta of SN~2007gr should be evident in the near-IR colour evolution at late times. Furthermore, the $H-K$ colour evolution tracks the evolution of the emission due to the first overtone of the carbon-monoxide molecule at $>$\,2.3 $\mu$m (see Fig. \ref{fig:5}  \S\, \ref{sec:CO}), which may be a precursor to dust condensation in the ejecta. Indeed, the $H-K$ colour curve of SN~2007gr tallies with the appearance and disappearance of the CO-feature between $\sim$ +\,70 to +\,200 days relative to $B$ maximum. Such an obvious signature of CO is not apparent in the $H-K$ colour curves of any of the other Ic SNe shown in Fig. \ref{fig:5}, with the possible exception of SN~1998bw which shows a mild upturn at about +\,60\,d, although the near-IR coverage for all objects other than SN~2007gr is admittedly lacking at epochs of interest. For SN~2007gr, given the relatively blue $H-K$ colour ($E_{H-K}$ is $\sim -1.2$ mag) at about +\,400\,d, we conclude that dust formation is unlikely to have occured in SN~2007gr by this epoch.

\subsection{The Bolometric (uvoir) Light Curves and Physical Parameters} \label{sec:Bolometric}

We constructed an Optical near-InfraRed (UVOIR) bolometric light curve by integrating the $UBVRIJHK$ broad band fluxes.  To derive the integrated SN luminosity, the magnitudes were corrected for reddening, converted to a flux density at the effective wavelength, and finally integrated using Simpson's rule.  The integrated flux was then converted to a luminosity using the distance modulus as assumed in \S\, \ref{sec:SN2007gr} i.e. $\mu$\,=\,29.84 $\pm$ 0.16. 

\begin{figure}[t]
	\begin{center}
	\includegraphics[width=0.5\textwidth, angle=0.0, trim=4mm 8mm 2mm 5mm, clip]{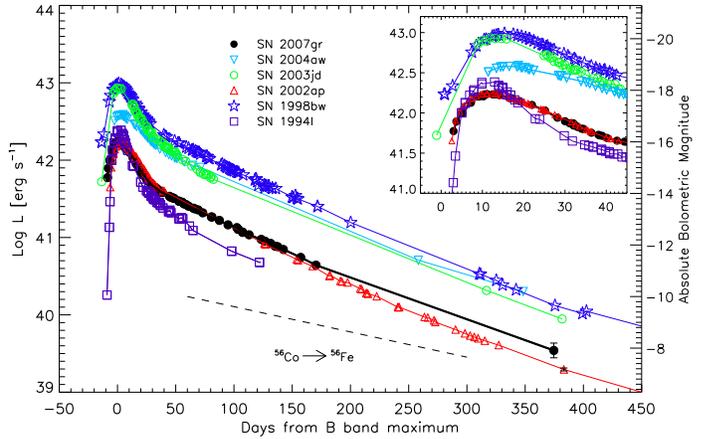}
	\caption{The uvoir bolometric light curve of SN~2007gr and other type Ic SNe.
                 The inset shows the differences in the widths of the uvoir lightcurves of this
                 sample of type Ic SNe.}
	\label{fig:Bol_LC}
	\end{center}
\end{figure}

The bolometric light curve of SN~2007gr is shown in Fig. \ref{fig:Bol_LC}.  For comparison, we have also included the bolometric light curves of the SNe Ic sample. Following \citet{Valenti08a}, we applied a toy-model to the bolometric light curve of SN~2007gr in order to constrain the physical parameters of the explosion (see Table \ref{tab:Ic_parameters}).  These include the ejected mass ($M_{\mathrm{ej}}$), the nickel mass ($M_{^{56}\mathrm{Ni}}$), and the kinetic energy ($\mathrm{E_{kin}}$). The procedure is based on a two-component analytical model to account for the photospheric and nebular phases of expansion.  During the photospheric phase (t $\leq$ 30\,d past explosion) homologous expansion of the ejecta, spherical symmetry, and the location of the radioactive $^{56}$Ni exclusively in the core is assumed \citep{Arnett82}. At late times (t $\geq$ 60\,d past explosion), the model includes the energy contribution from the $^{56}$Ni$\rightarrow$$^{56}$Co$\rightarrow$$^{56}$Fe decay, following the work of \citet{Sutherland84} and \citet{Cappellaro97}.  In the nebular phase the ejecta are optically thin, so that the incomplete trapping of $\gamma$-rays and positrons has to be accounted for \citep{Clocchiatti97}. 

The modelling of the bolometric light curve suggests that the ejected mass $M$$_{ej}$ of SN~2007gr (2\,-\,3.5$M_{\sun}$) is similar to SN~2002ap but the kinetic energy $E_{kin}$ is considerably lower (by a factor of $\sim$3).  However, we do note that the parameters for SN~2002ap (see Table \ref{tab:Ic_parameters}) have been derived via a more involved modelling procedure \citep[e.g.][]{Mazzali02,Foley03}.

A comparison of the early quasi-bolometric light curves shows that SN~2007gr appears to be most similar to SN~2002ap in terms of peak luminosity and width. As alluded to previously, the similarity in peak luminosity ($M$$_{\mathrm{Bol}}=-16.9)$ implies that these SNe produced similar quantities of $^{56}$Ni ($\sim0.07-0.1\,M_\odot$).  We estimate that the quantity of $^{56}$Ni ejected in SN~2007gr is 0.076 $\pm$ 0.020, similar to SNe~ 2002ap and 1994I. Interestingly, SN~2007gr also resembles SN~2003jd in the shape of the early bolometric light curve, although at a much lower luminosity, while the rise to peak luminosity is comparable to SN~1994I.  The rise time of stripped envelope SNe depends both on the radius of the progenitor prior to explosion and on the quantity of ejected mass \citep{Arnett96}.  
For reference, we list the rise times of a sample of Ib/c SNe in Table \ref{tab:Ic_parameters} 
collected from the literature, but note that for all objects other than SN~2007gr, these are 
observationally poorly constrained, and the reality of any differences in the rise times cannot 
at this point be directly attributed to physical differences between the progenitor stars.

SNe~2007gr and 2002ap have similar peak widths which are evidently broader than that of SN~1994I (see Figure \ref{fig:Bol_LC} inset). Such broad peaks are usually attributed to more massive SNe Ic progenitors \citep[e.g.][]{Arnett82,Mazzali01,Foley03}.  Since the peak widths indicate the rate of energy diffusion out of the core, this suggests that the progenitors of SNe~2007gr and 2002ap had relatively large masses.  We also expect the mass ejected in these two events to be larger compared to SN~1994I but less than for SNe~1998bw and 2004aw.

The bolometric light curve of SN~2007gr indicates a late-time (120 to 380 day) decline of 
$\gamma$ = 0.0142 $\pm$ 0.0002 mag d$^{-1}$.  These decline rates are comparable to those of the broad-lined SNe~1998bw and 2002ap which are $\gamma$ = 0.0156 $\pm$ 0.0001 mag d$^{-1}$ and $\gamma$ = 0.0162 $\pm$ 0.0003 mag d$^{-1}$, respectively.  The decline rates are faster than expected from the decay of radioactive $^{56}$Co (i.e. 0.0098 mag d$^{-1}$). If the decay of $^{56}$Co is responsible for powering the late-time light curve of SN~2007gr, then this suggests that the steeper decline is due to a decreasing $\gamma$-ray trapping in the SN envelope with time.

In Table \ref{tab:Ic_parameters}, we compare SN~2007gr to other type Ic SNe which show a wide range of physical properties at the explosion epoch.  We can see that SN~2007gr is most similar to SNe~2002ap and 1994I in terms of the amount of $^{56}$Ni synthesised (see \S\,\ref{sec:Bolometric}).   It can also be noted that SN~2007gr appears to be similar to SN~1994I with respect to the ratio of kinetic energy to ejected mass.  Thus SN~2007gr may be considered as the link between broad-lined events like SN~2002ap and low energy events like SN~1994I.  We note however, that the parameters listed in Table \ref{tab:Ic_parameters} have been derived using a variety of methods.

\section{Spectroscopy} \label{sec:Spectroscopy}
\subsection{Optical spectra} \label{sec:Optical_spectra}

\begin{figure*}[h]
\begin{center}
\includegraphics[width=0.85\textwidth,angle=0.0, trim=8mm 0mm 0mm 4mm, clip]{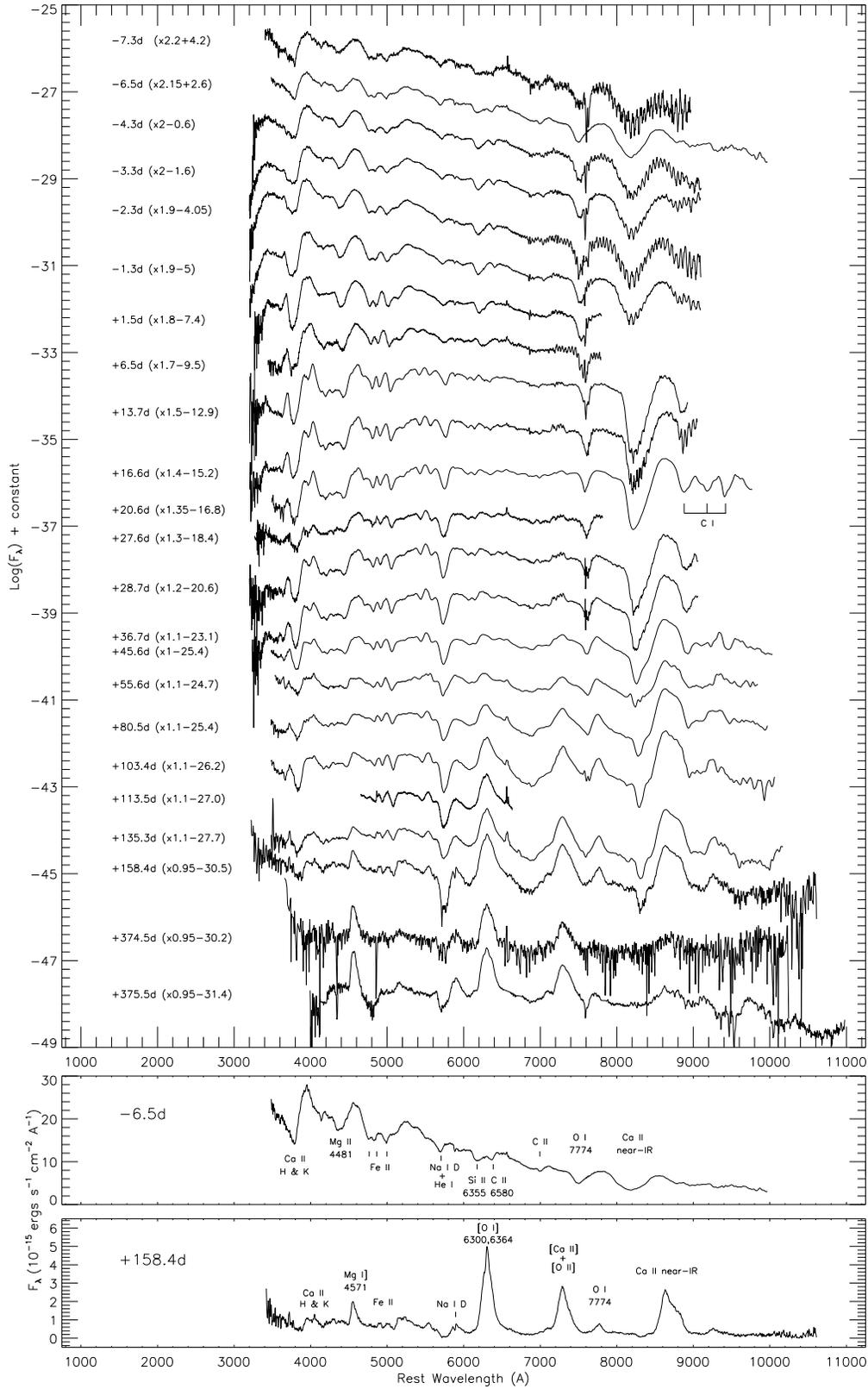}
\caption{Evolution of the optical spectra of SN~2007gr.  The spectra have been corrected for 
reddening, $E(B-V)$$_{tot}$\,=\,0.092, and for the host galaxy redshift of NGC 1058, $z$\,=\,0.001728.  The spectra have also been magnified and displaced vertically for clarity by the numbers shown in brackets.  Identifications are shown in the spectra of SN~2007gr at $-$6.5 and 158.4 days relative to $B$ maximum.}
\label{fig:Optical_spectra}
\end{center}
\end{figure*}

The spectroscopic evolution of SN~2007gr, from 7.3 days prior to maximum light 
to +\,375.5 days after, is densely covered and is shown in Fig. \ref{fig:Optical_spectra}.  The line identifications are based on those by \citet{Valenti08a}.

The early spectra exhibit Ca II H$\&$K, Mg II $\lambda$ 4481, and Na~I~D $\lambda$$\lambda$ 5891, 5897 which are responsible for the P-Cygni absorption features observed at 3810-3850\,$\AA$, 4390-4490\,$\AA$, and 5720-5760\,$\AA$, respectively.  The Si II 6355\,$\AA$ absorption can also be identified in the early spectra, however it fades around two weeks after maximum light.  

\citet{Chornock07} classified SN~2007gr as a type Ib/c from a spectrum obtained the night following discovery while waiting for follow-up observations.  Eventually, no strong evidence for 6678 and 7065\,$\AA$ He features which are typically in type Ib SNe $\sim$ 7\,days after maximum light were found. Consequently, SN~2007gr was classified as a type Ic.  \citet{Valenti08a} studied the feature at 6450\,$\AA$, which may be misidentified as 6678\,$\AA$ He I in early-time spectra, and suggest that a more plausible identification is C II\,6580 $\AA$  at a velocity of $\sim$11,000 \kms.  Indeed this feature disappears around maximum together with another feature at 7000\,$\AA$ attributable to C II 7235\,$\AA$, thereby providing support for the Ic classification. 

Moreover, almost two weeks after $B$ maximum, one of the strongest optical transitions of the C I ion (9095\,$\AA$) appears, reaching velocities of $\sim$8000 \kms.  The spectrum at around three weeks display two additional C I absorptions (9094, 9406\,$\AA$) due to the extended coverage at longer wavelengths \citep{Valenti08a} (see section \ref{sec:Optical_comparison} and \label{sec:nir_analysis} for further discussions).

The first ten spectra, extending from one week before $B$ maximum to one week after, are characterised by an initially blue continuum which reddens significantly with time. Following maximum light, the Ca II near-IR triplet and the Na I D lines become very prominent in the spectra.  Other absorption features include Fe II (4924, 5018, 5169\,$\AA$) and Sc II (5552\,$\AA$).  Approximately one week after maximum, the absorption troughs centred at 4200\,$\AA$ and 4800\,$\AA$ show the $\emph{W}$- shaped profile that is characteristic of many type Ic SNe at a comparable epoch. By this time, the flux in the bluer bands is significantly reduced which is consistent with the redder $U-B$ and $B-V$ colours (Fig. \ref{fig:5}).

The spectral evolution of SN~2007gr is relatively fast with the disappearance of Si II 6355\,${\AA}$ already two weeks after maximum.  The relatively early appearance of [Ca II] + [O II] $\sim$\,7300\,$\AA$ and [O I] $\lambda$$\lambda$ 6300, 6364 at about +\,55\,d represents the onset of the nebular phase. The absorption features in the spectroscopic evolution of SN~2007gr tend to drift towards the red with time, as expected.  This is a result of the formation of lines occuring in slower, deeper layers of the ejected material.  

\clearpage

The late-time spectra are dominated by the emission of forbidden lines such as Mg I] 4571\,$\AA$, [O I] $\lambda$$\lambda$ 6300, 6364, [Ca II] $\lambda$$\lambda$ 7291, 7323, and the permitted transitions of Na I D, O~I~7774, and the possible blend of C I (8727\,$\AA$) and the Ca II near-IR triplet.  

The optical spectra do not reveal any clear indication for the presence of He I lines.  If 
He I 5876\,$\AA$ was present it would almost certainly be contaminated by Na I D. Other 
lines of He I in this region (e.g. 4471, 7065\,$\AA$) would also be blended with other strong features at these wavelengths.  Since the optical spectra alone are insufficient to detect small amounts of He I, in section \ref{sec:nir_analysis} we examine the near-IR spectra of SN~2007gr. The He I 10830\,$\AA$ line is expected to be the strongest He I transition \citep{Swartz93} and therefore should be highly sensitive to very small quantities of helium \citep{Wheeler93,Baron96}.

\subsection{Optical Spectra Comparison}
\label{sec:Optical_comparison}

\begin{figure*}[h]
	\centering
		\includegraphics[width=0.95\textwidth, trim=0mm -2mm 0mm 0mm]{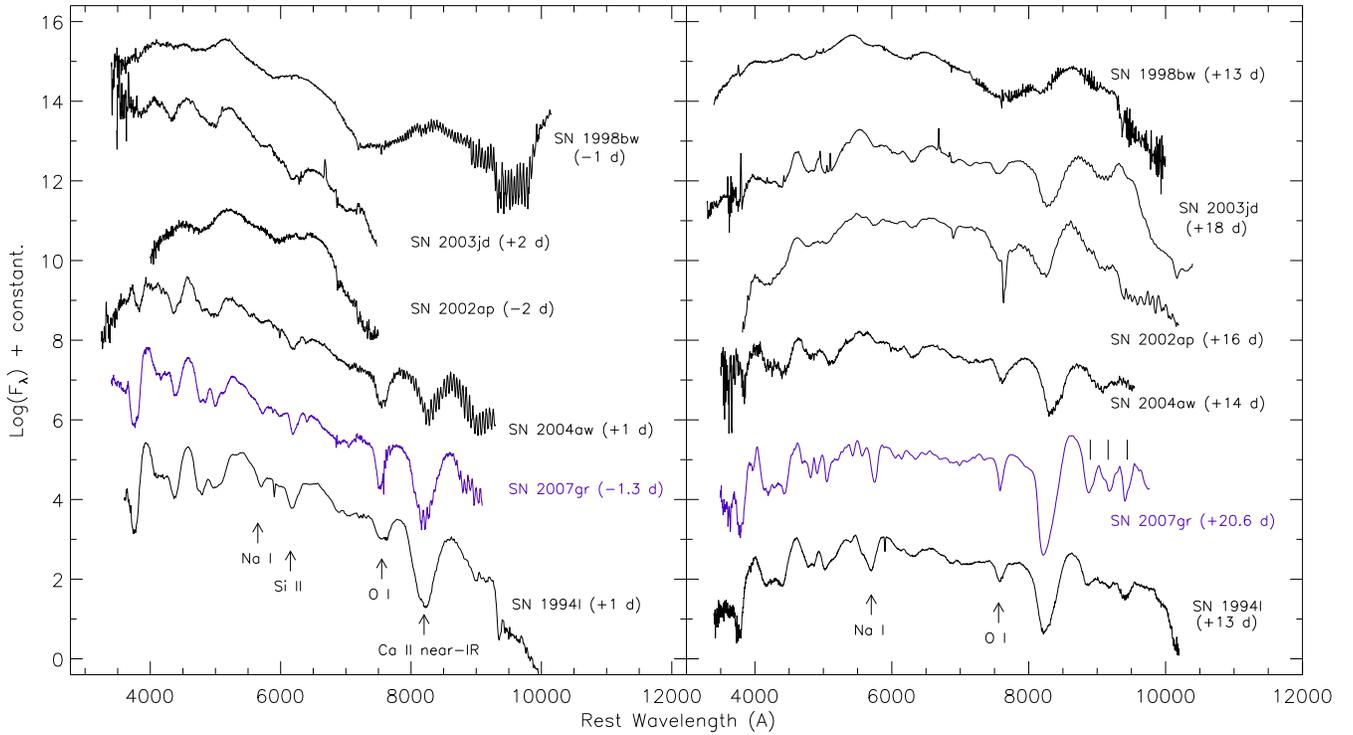}
	\caption{Comparison of spectra of SNe close to $B$-band maximum (left) and two-three weeks post maximum light (right). All spectra are de-reddened and redshifted assuming the values reported in Table \ref{tab:Ic_parameters}.  Typical absorption lines are identified.  The three solid lines in the right-hnd panel next to SN~2007gr indicate the positions of the C I lines. SN~1994I also shows the C I features, although these are not as strong as in SN~2007gr, while the other SNe show a broad dip in this region; these variations are probably attributable to the differences in the respective ejecta velocities of the SNe shown here, with the C I features being smeared out in those with the highest velocities.  }
	\label{fig:Optical_spectra_Ic_1}
\end{figure*}

\begin{figure*}[h]
	\centering
		\includegraphics[width=0.95\textwidth, trim=0mm -2mm 0mm 0mm ]{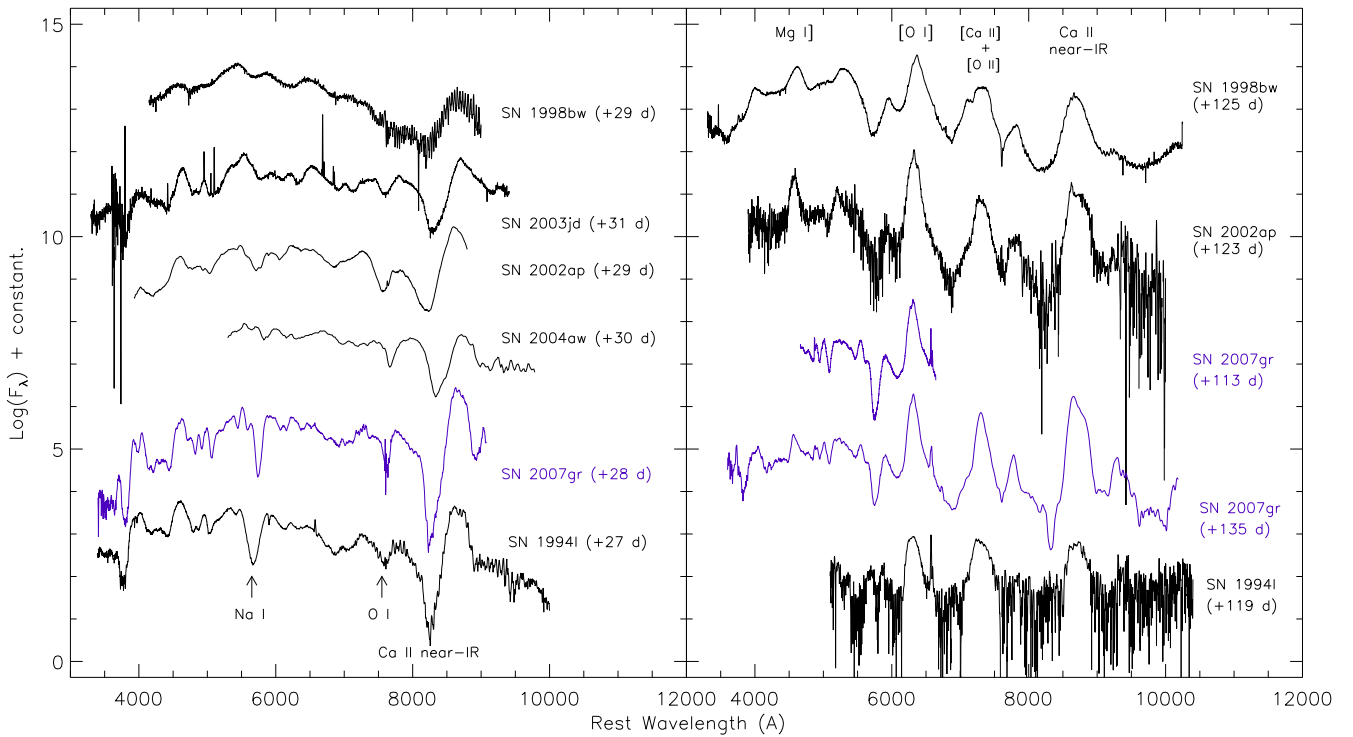}
	\caption{Comparison of spectra of SNe about one month (left) and four months (right) after $B$ maximum light.  The P-Cygni absorptions of Na I D ($\lambda$$\lambda$ 5891, 5897), O I ($\lambda$ 7774) and the Ca II near-IR triplet are clearly marked in the left panel.  Typical nebular emission lines are indicated in the right panel. }
	\label{fig:Optical_spectra_Ic_2}
\end{figure*}

In Figures \ref{fig:Optical_spectra_Ic_1} \& \ref{fig:Optical_spectra_Ic_2}, we present early and 
late-time spectral comparisons of SN~2007gr with the SNe Ic sample: SN~1998bw \citep{Sollerman00,Mazzali01}, SN~2003jd \citep{Valenti08b}, SN~2002ap \citep{Gal-Yam02,Kinugasa02,Foley03}, SN~2004aw \citep{Taubenberger06} and SN~1994I \citep{Filippenko95b,Sauer06}.  All of the spectra have been ordered according to their explosion energies (top spectra resulting from the most energetic) and corrected for reddening and redshift using the values listed in Table \ref{tab:Ic_parameters}.  We have previously emphasised the remarkable similarity in light curve morphology and evolution between SNe~2007gr and 2002ap. We will see that this similarity does not extend to the spectral line profiles and ejecta expansion velocities.

The near-maximum spectrum of SN~2007gr resembles that of SN~1994I on day +1 in terms of the presence of all major features.  The main differences between these events at this epoch appears to be the deeper absorption of O I 7774 $\AA$ relative to Na I 5896 $\AA$ and the narrower spectral widths in SN~2007gr.  In contrast, the spectra of SNe~1998bw and 2002ap exhibit very broad features as a result of line blending from high-velocity ejected material. The narrow lines of SNe~2007gr and 1994I allow for a more secure line-identification and a better separation of spectral features.  This is especially true for the C I absorptions (8335, 9094, 9406 $\AA$) which are evident in SN~2007gr at +\,20.6\,d and are also detected in SN~1994I, although at weaker intensities.  For SNe~2003jd and 2004aw, it is possible that the individual C I lines may be disguised as a single absorption blend at $\sim$\,9000\,$\AA$.

SNe~2007gr and 1994I also have better developed P Cygni profiles for the Na I D and Ca II near-IR triplet lines at only two weeks past maximum as a consequence of their rapid evolution.  The comparatively slowly evolving SNe~1998bw, 2004aw, and 2003jd, show no sign of these features until $\sim$2 months.

Four months after $B$ maximum, the spectra of SN~1994I are the most evolved and show only forbidden line emissions.  The absence of the photospheric O I 7774\,$\AA$ emission line also indicates that the evolution from the photospheric to nebular phase is faster than in SNe~1998bw, 2002ap and 2007gr.  Even at this nebular phase, the spectral lines of SN~2007gr remain narrower than the other type Ic SNe presented, indicating the slower expansion velocities of the ejecta.  The nebular phase spectra all show [O I] $\lambda$$\lambda$ 6300, 6364 as one of their strongest features, complemented with lines of [Ca II] $\lambda$$\lambda$  7291, 7323 / [O II] $\lambda$$\lambda$ 7300, 7330, and the near-IR Ca II triplet.  The largely isolated nature of the [O I] doublet is found to be narrower in SN~2007gr compared to all the SNe Ic presented including SN~1994I by $\sim$\,1000 \kms.

The nebular spectra of SN~2002ap, 1998bw and 2007gr have strong [O I] $\lambda$$\lambda$ 6300, 6364 and Mg I] $\lambda$ 4571 emissions. The ratio of Mg I] to [O I] may provide some information on the degree of stripping of the outermost layers experienced by the progenitor star, with a higher Mg I]/[O I] ratio implying a greater degree of stripping as more of the O-Ne-Mg layer is exposed \citep{Foley03}.  Additionally, a higher Mg I]/[O I] ratio at later epochs may simply imply that less of the outer C/O layer is observed due to the transparency of the ejecta. 
We note that the [O I] and Mg I] lines may well be formed via different mechanisms, with  [O I] being collisionally 
excited while Mg I] arising predominantly due to recombination as suggested by \citet{Kozma98}. This could influence the line ratios that 
we measure in different SNe. Detailed modelling of the excitation mechanisms for these lines is beyond the scope of this work, so we
merely flag this caveat, and use the line ratios as a proxy for the composition.

\begin{figure*}[h]
	\centering
		\includegraphics[width=0.8\textwidth,angle=0.0]{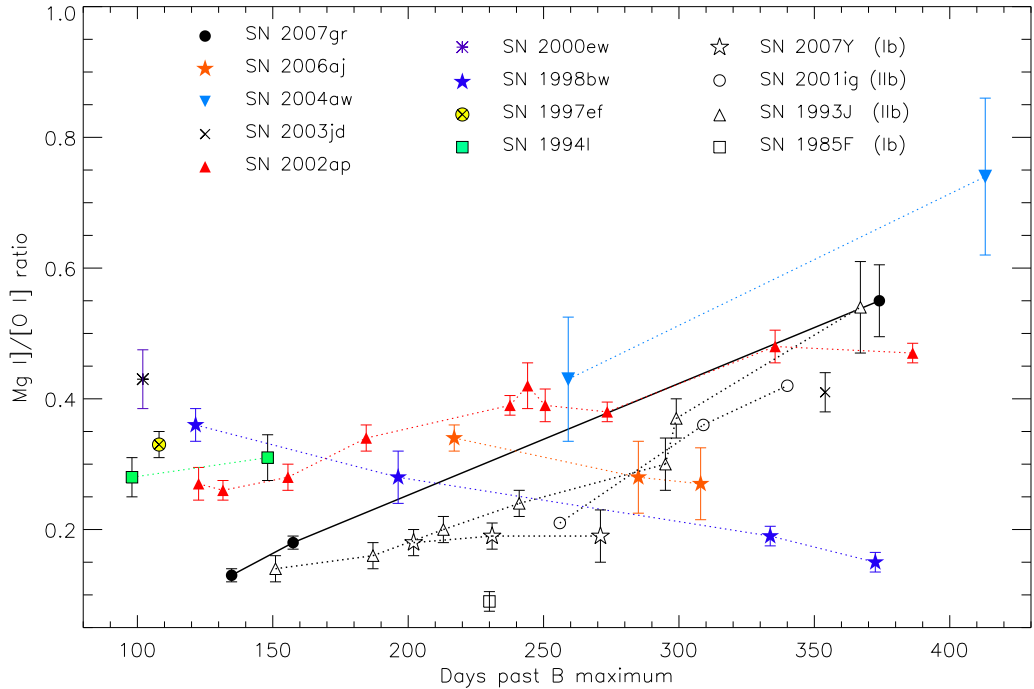}
	\caption{The Mg I]/[O I] ratio as a function of time for a sample of SNe.
                 All measurements were performed as described in section \ref{sec:Optical_comparison}, with the exception
                 of type IIb SN~2001ig, where the measurements were taken from \citet{Silverman09}.
                 All other Ic SNe spectra are from references listed in Table \ref{tab:Ic_parameters} with the exception of SN~2006aj \citep{Mazzali07,Taubenberger09} and 2000ew (Valenti et al. in prep).  The spectra of the type Ib \object{SNe~2007Y} \citep{Stritzinger09} and 1985F \citep{Filippenko86}, and the type IIb SN~1993J \citep{Fransson05} have also been used.  They were corrected for extinction before the measurements were made.}
	\label{fig:10}
\end{figure*}

In Figure \ref{fig:10} we show the evolution of the Mg I]/[O I] ratio for a sample of CCSNe during the nebular period +\,100 to +\,400\,d.  The Mg I]/[O I] ratio was calculated by first fitting a cubic spline to a local continuum. The continuum was then subtracted so that the integrated flux in the emission lines could be estimated. All of the SNe in Figure \ref{fig:10} that have more than one epoch of measurement available, show a trend of increasing Mg I]/[O I] with time, with the marked exceptions of SNe~1998bw and \object{2006aj}. Interestingly, these two SNe are the only SNe in this sample that were associated with long-duration GRBs. For the SNe for which only a single measurement was possible (\object{SNe~1985F}, \object{1997ef}, \object{2000ew}, 2003jd), the Mg I]/[O I] ratio was found to lie within the bounds spanned by other SNe of this type, and are included in Fig. \ref{fig:10} for completeness.

The increasing Mg I]/[O I] trend observed for the majority of the SNe sample may simply be due to an abundance effect, as hypothesised by \citet{Foley03}.  In symmetric or quasi-symmetric explosions, the onion-structure of the progenitor star is almost preserved.  This is also true in non-symmetric explosions viewed equatorially.  In these cases as time goes by we observe deeper and deeper into the ejecta reaching the O-Ne-Mg shell, and so the Mg I]/[O I] is expected to increase.

At the beginning of the nebular phase (+\,120 to +\,170\,d), SN~2007gr has lower Mg I]/[O I] ratios than the other SNe Ic. The lower Mg I]/[O I] ratio of SN~2007gr at these early epochs may suggest that we are still observing the highly concentrated C/O layer of the outer ejecta.  This may support the idea that SN~2007gr originated from a carbon-rich progenitor (see \S\,\ref{sec:nir_analysis}).  Another possible reason for the lower Mg I]/[O I] ratio may be due to the inefficient burning of the large quantities of carbon in SN~2007gr, thus producing only small quantities of Mg. Conversely, the larger Mg I] to [O I] ratios of SN~2002ap may imply that the progenitor experienced greater stripping of the outer carbon layer, since at an epoch similar to SN~2007gr we are already probing deeper into the core of the progenitor, revealing the O-Ne-Mg layer \citep[e.g.][]{Foley03}.

The data also show that SNe~2007gr and 2002ap exhibit larger Mg I] to [O I] ratios than the type Ib SNe~1985F \citep{Filippenko86} and 2007Y \citep{Stritzinger09}.  In this figure, we have also included the ratios of the type IIb \object{SNe~2001ig} \citep{Silverman09} and \object{1993J} \citep{Fransson05}.  The Mg I]/[O I] ratios of the type IIb SNe are also consistently lower than those of the type Ic SNe~2007gr, 2002ap and 2004aw at the same epochs.

\clearpage

\begin{figure*}[t]
	\centering
	  \includegraphics[width=0.8\textwidth, trim=0mm -5mm 0mm 0mm]{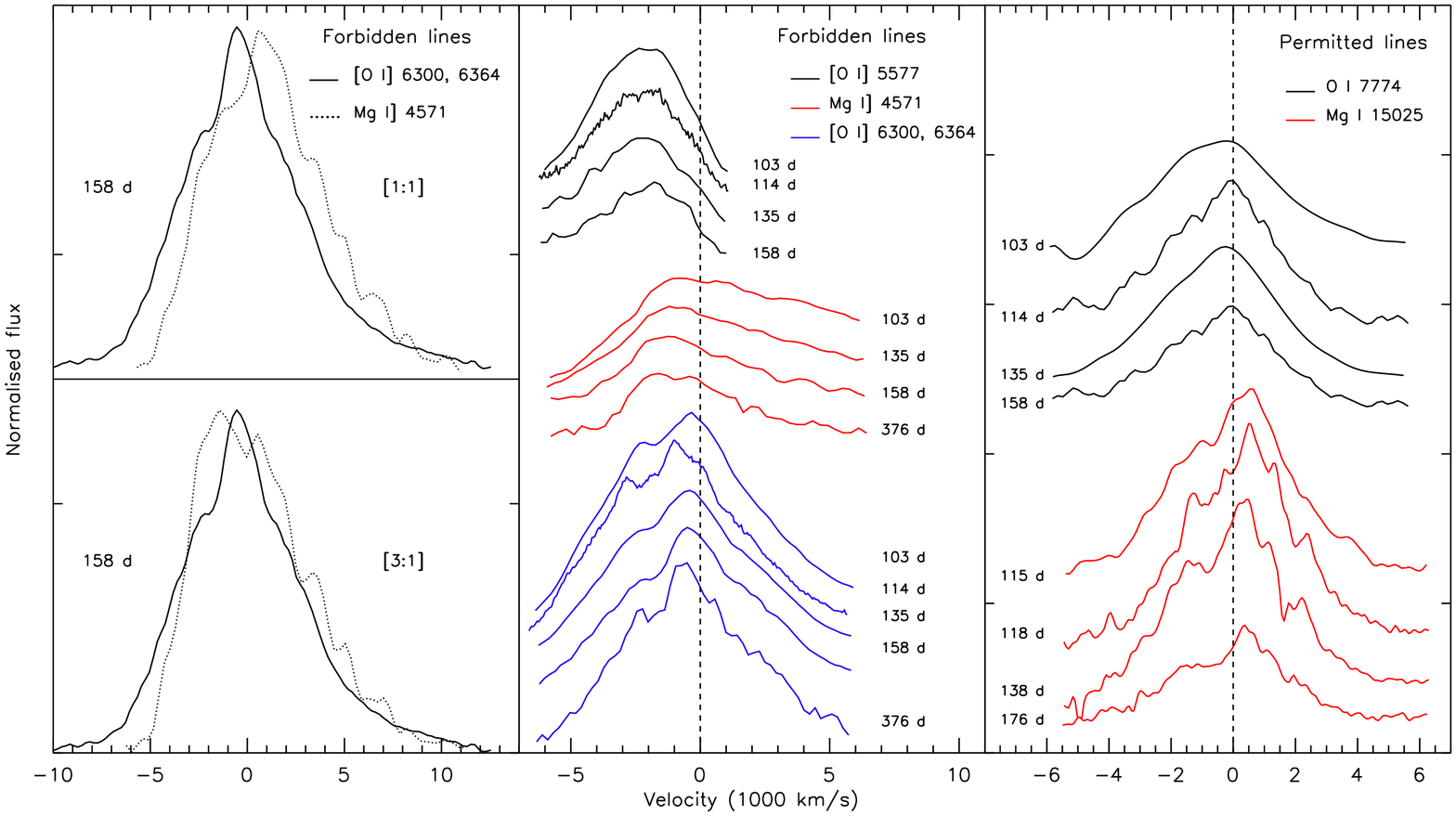}
	\caption{$\it{Left:}$ The profiles of [O I] 6300, 6364 and Mg I] 4571 at 158\,d past $B$ maximum. The profile of [O I] is plotted in velocity space using a rest wavelength of 6315\,$\AA$, as obtained via gaussian fitting from the modified profile of [O I] in the normal and low-density limits, 1:1, 3:1.  The Mg I] profile includes the contribution of an artificial component with an intensity ratio of 0.33 relative to the real peak and the same velocity offset as found in the oxygen doublet.  The peaks of the Mg I] profiles have been scaled to match the peak of the oxygen doublet at +\,158\,d.  $\it{Centre:}$ Evolution of forbidden [O I] 5577, Mg I] 4571 and [O I] 6300, 6364.  $\it{Right:}$  Evolution of the profiles of the permitted lines of O I 7774\,$\AA$ and Mg I 15025\,$\AA$.  The vertical line in the panels correspond to zero velocity i.e. the rest wavelengths of the line profiles.  }
	\label{fig:asymmetry}
\end{figure*}
 \clearpage

The general trend from our sample supports the idea that the progenitors of type Ic SNe experience greater envelope stripping than those of type Ib and IIb. 

The decreasing Mg I]/[O I] trend observed for SNe~1998bw and 2006aj may be explained by the rapid rotation of the GRB-progenitors which result in greater mixing (via meridional circulation), and probably induce some degree of asymmetry. A complete mixing of the onion structure is expected in non-symmetric explosions viewed along the polar axis (along the jet).  
However, it may be that because of the efficiency of this mixing mechanism, the early Mg I]/[O I] ratios will be close to the average values observed in other Type Ic SNe (e.g. first point of SN~1998bw).  As time progresses, the quantity of Mg will diminish and since O is more abundant, its lines will dominate over Mg lines. In this simple view, the SNe which show the trend of SNe~1998bw and 2006aj could be jet-like explosions or explosions where the ejecta has been fully mixed.  An alternative interpretation may be that the ionisation equilibrium evolves with time for Mg and O (which have very different ionisation potentials) simply because of decreasing temperature \citep[cf.][]{Meyerott80}.  This may well be a contributing
factor that might account for the observed differences.

Recent studies of nebular spectra suggest that around half of all SNe Ib/c are strongly 
aspherical while the remainder appear to have moderate signatures of asphericity 
\citep{Maeda08,Modjaz08,Taubenberger09,Milisavljevic09}. As previously mentioned, [O I] $\lambda$$\lambda$\,6300, 6364 and Mg I] $\lambda$\,4571 are two of the most dominant features in the late-time spectra of stripped-envelope SNe.  As a consequence of their relatively isolated nature and higher abundances to other nebular lines, an analysis of the emission profiles permits one to study the ejecta geometry \citep{Mazzali05}. In particular, one might expect Mg and O to have a similar spatial distribution within the ejecta \citep[e.g.][]{Maeda06} and thus the profiles of these emission lines should also be similar. 

Unfortunately, a direct comparison of the Mg I] and [O I] lines is not possible since the [O I] doublet arises from two forbidden transitions which share the same upper level ($^{3}$P$_{1,2}$ - $^{1}$D$_{2}$).  The relative intensity of the $\lambda$\,6300 and $\lambda$\,6364 lines will depend on the O I density and may vary from 1:1 to the low-density limit of 3:1 \citep{Spyromilio94}.  \citet{Taubenberger09} suggested that the density in stripped envelope SNe should, after 100\,d, have decreased enough to fix the line intensity ratio to 3:1, but recently \citet{Milisavljevic09} suggested a ratio 1:1 to explain the O I $\lambda$$\lambda$\,6300, 6364 profiles in 4 stripped envelope SNe.  In order to investigate which line intensity ratio is more appropriate for SN~2007gr, we have displayed the profiles of [O I] $\lambda$$\lambda$ 6300, 6364 and Mg I] $\lambda$ 4571 in Figure \ref{fig:asymmetry} (left panel) at 158\,d after maximum. Since the profiles of Mg I] (+ an artificial second component with a line intensity ratio of 0.33 and the same velocity offset as found in the oxygen doublet) and [O I] are similar, this tells us that the ratio between O I 6300 and 6364 is close to 3:1.  We also performed a comparison between [O I] $\lambda$$\lambda$\,6300, 6364 and [O I] $\lambda$\,5577 (+ an artificial second component with ratio 0.33, 0.66, 1.00) without obtaining a good match.  The [O I] emission profile contains one peak positioned at a blueshift velocity of $\sim$\,-1500 \kms and the other skewed at $\sim$\,-800 \kms. \citet{Taubenberger09} found an average blueshift of $\sim$\,1000\,\kms\ for the [O I] $\lambda$\,6300 line centroids of a large sample of Type Ib/c SNe which is also consistent with SN~2007gr.  The authors suggest that the observed effect is probably related to the ejecta geometry and may be a signature of a one-sided explosion, or, more likely, residual opacity in the inner ejecta. 

We also present the evolution of the permitted and forbidden lines of O and Mg at phases $>$ +\,100\,d (centre and right panels of Figure \ref{fig:asymmetry}).  The figures show that the permitted lines of O I $\lambda$ 7774 and Mg~I~$\lambda$ 15025 behave very differently than the forbidden lines of [O I] $\lambda$ 5577 and Mg I] $\lambda$ 4571.  The emission peaks of the permitted profiles are found to be almost centered at their rest wavelengths while those of the forbidden lines are strongly blueshifted, with the effect becoming progressively weaker with time.  In addition, the [O I] profile at 5577 $\AA$ appears to be even more strongly blue-shifted compared with other forbidden lines.

The observed trend of blueshifted forbidden line centroids compared to the central permitted lines suggest that they appear in different regions of the ejecta since we know that the forbidden lines emerge from a lower-density medium.  
The systematic effective blue-shift of the forbidden lines may well be due to the residual opacity of the intervening material such
that the photons emitted on the rear side of the SN are absorbed along the line of sight.

\citet{Taubenberger09} and \citet{Milisavljevic09} provide alternative explanations for the observed trend, including dust formation and contamination from other ions.  However, in section \ref{sec:CO} we claim that dust has not yet formed in the ejecta of SN~2007gr.  Additionally, any dust that is present should actually result in a line blueshift which increases with time, not the contrary.  Contamination due to other emission lines may also be neglected due to the similarity of the [O I] and Mg I] profiles, while contamination on the [O I] $\lambda$\,5577 cannot be excluded.

For SN~2007gr, we note that \citet{Tanaka08} report non-zero polarization across the Ca near-IR triplet at $\sim$21\,d post-maximum. They concluded that their observations could be interpreted either in the context of a bipolar explosion with an oblate photosphere viewed from a mildly off-axis line-of-sight, or a spherically symmetric photosphere with a clumpy Ca II distribution. Out to about 400\,d, the optical and near-IR spectra, show no evidence of double-peaked profiles as would be expected for a significantly aspherical explosion. From our data, we conclude that the explosion of SN~2007gr could not have been more than mildly aspherical.

\subsection{Near-infrared spectra}
\label{sec:nir_analysis}

\begin{figure*}[t]
\centering
\includegraphics[width=1.0\textwidth, trim=5mm 17mm 5mm 0mm]{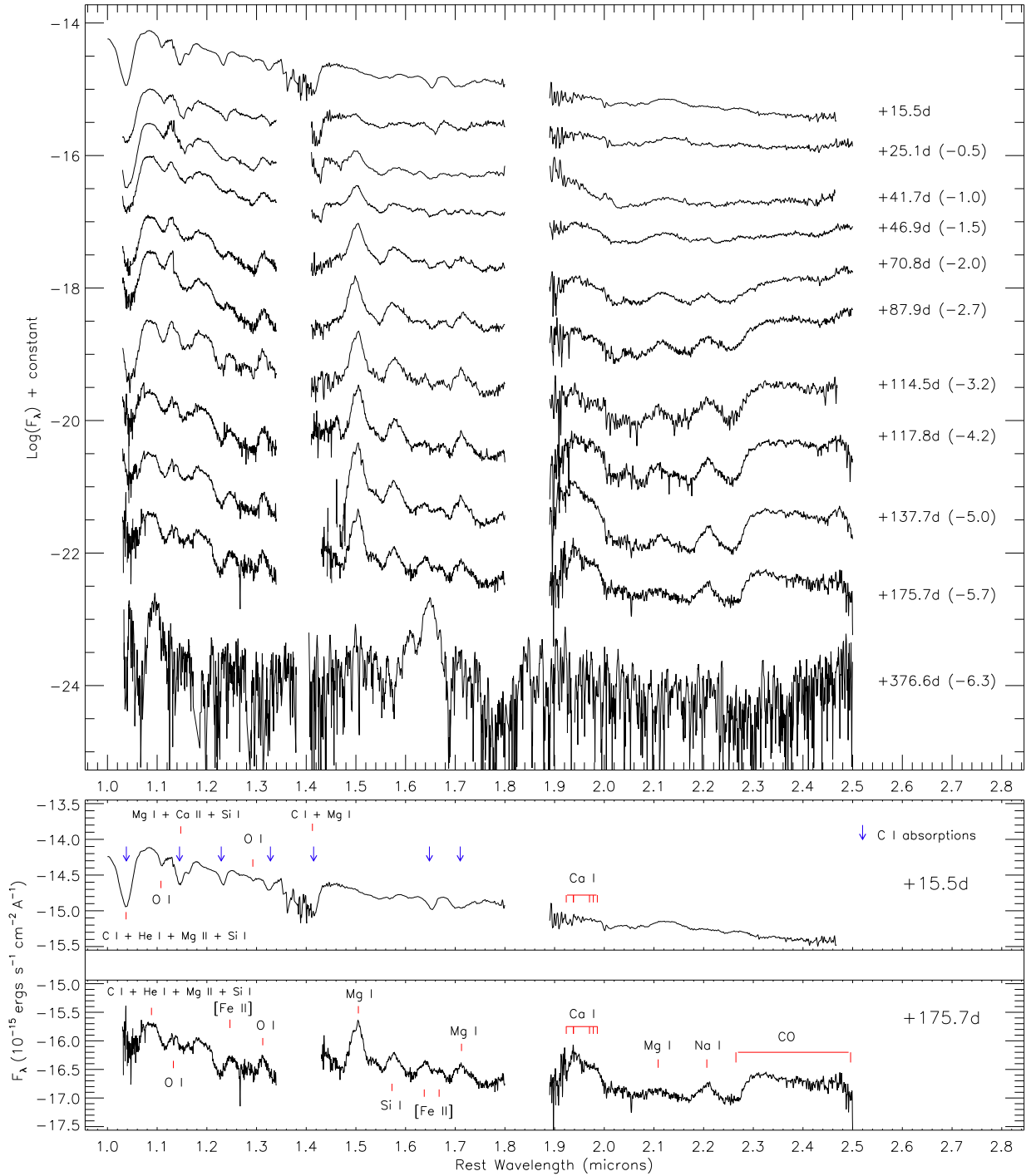}
\caption{The near-infrared spectral evolution of SN~2007gr. The spectra are displayed in the rest frame of NGC~1058. To increase the temporal coverage we include the spectrum obtained by \citet{Valenti08a} on day +\,15. The +\,376.6\,d spectrum has been smoothed with a box size of 5 pixels. The spectra have been displaced vertically for clarity by the numbers shown in brackets. The rest wavelengths of possibly identified lines are also included. \label{fig:NIR_spectra}}
\end{figure*}

All near-infrared spectra of SN~2007gr are presented in Fig. \ref{fig:NIR_spectra}.  At early epochs, the shape of the NIR continuum is nicely approximated by a black body continuum.  
The earliest spectra are almost featureless with few lines due to intermediate mass elements 
such as  Mg I, O I, Na I.  After maximum light, lines due to the iron group elements begin to appear. 

One of the most prominent features in the near-IR spectra is an absorption at $\sim1.04\,\mu$m. The origin of this feature has been discussed widely by several authors, especially in the 
context of type Ia SNe, with \citet{wheeler:98} favouring Mg II and \citet{meikle:96} suggesting a blend of Mg II and He I 1.083\,$\mu$m; the latter is a line that is easily excited in a variety of astrophysical contexts.  In fact, other elements such as C, O, Si, and Fe may also contribute to features in this region.

 If He I were indeed responsible for the absorption, it would only be detected if the lower level of He becomes highly populated in the presence of high energy, non-thermal radiation.  The source of this radiation would be the $\gamma$-rays produced in the radioactive decay of $^{56}$Ni to $^{56}$Co. The lower level population may then absorb photospheric photons of $\sim$1 micron yielding the P Cygni profile.  
 
\clearpage

\citet{Mazzali98} were able to reproduce the 1.05 micron feature in the Type Ia SN~1994D using just 0.01 M$_\odot$ of helium, implying that only small quantities are needed to account for this feature. 
Thus the apparent lack of strong He lines in type Ic SN spectra may imply a low abundance of He I and/or insufficient mixing of $^{56}$Ni into the outermost layers of the ejecta.

If He I were a major contributor to the 1.04\,$\mu$m feature in SN~2007gr, this would imply an average expansion velocity of $\sim$13,000\,$\kms$ at 15.5\,d which later declines to $\sim$10000\,$\kms$ by 114.5\,d. In spite of the fact that SN~2007gr is a narrow-lined Ic, so that the identification of lines should be facilitated, the optical region is plagued by blends, especially in the regions where He I lines would lie e.g. the He I 3888\,$\AA$ line lies close to the Ca II H \& K lines and the He I 5875\,$\AA$ lies close to the Na I D. The possible presence of the He I line at 5875\,$\AA$ is suggested in Section 5.4 to account for the unusual velocity distribution of the Na I D feature.  In the near-IR, one might expect to detect He I 2.0581\,$\mu$m.  However, we see no evidence of any feature at this wavelength in our entire near-IR spectral series.

If He I were a major contributor to the 1.04\,$\mu$m feature in SN~2007gr, this would imply an average expansion velocity of $\sim$13,000\,$\kms$ at 15.5\,d which later declines to $\sim$10000\,$\kms$ by 114.5\,d. In spite of the fact that SN~2007gr is a narrow-lined Ic, so that the identification of lines should be facilitated, the optical region is plagued by blends, especially in the regions where He I lines would lie e.g. the He I 3888\,$\AA$ line lies close to the Ca II H \& K lines and the He I 5875\,$\AA$ lies close to the Na I D. The possible presence of the He I line at 5875\,$\AA$ is suggested in Section 5.4 to account for the unusual velocity distribution of the Na I D feature.  In the near-IR, one might expect to detect He I 2.0581\,$\mu$m.  However, we see no evidence of any feature at this wavelength in our entire near-IR spectral series.

Although He I 2.0581\,$\mu$m is not visible in SN~2007gr, we cannot fully disregard the contribution of helium to the 1.04 \,$\mu$m absorption. The longer decay time from the 2s3S level of He~I 1.0830\,$\mu$m means that this line can behave much like a low-energy, ground state transition, unlike the faster ($\times$ 4.5x10$^{5}$\,s) decay time of He I 2.0581\,$\mu$m, and so will be easier to excite.  Thus the absence of the He I 2.0581\,$\mu$m line does not necessarily mean that the 1.0830\,$\mu$m line does not contribute to the 1.04\,$\mu$m feature. This cannot be quantified without detailed modelling which is beyond the scope of this paper.  Yet, we can conclude that if He is present in SN~2007gr, the amount must be small.

\citet{Valenti08a} suggested that the 1.04\,$\mu$m feature in SN~2007gr may be attributed primarily to the C I ion (rest wavelength at 1.0695\,$\mu$m). They found that earlier (optical) spectra showed evidence for C II features at very early epochs ($\sim-8$\,d) which fade rapidly, to be replaced by lines due to C I.  

From an initial modelling effort, they obtained a fairly reasonable match to their optical-near-IR 
spectrum at day +\,15, also reproducing several other observed C I lines. However, even with a model enhanced in C I, the entire 1.04\,$\mu$m feature, and the blue wing in particular, could not be matched. They suggested that this could be due to the contribution of one or more other ions e.g. Mg II, Si II. Indeed \citet{Millard99} also had difficulty in accounting for this infrared absorption feature in SN~1994I with He I alone, suggesting that it could be a blend of He I and C I lines.     

Together with the 1.04\,$\mu$m feature, the nebular spectra of the $J$ band reveal emission lines of Fe II, Si II, Mg I, Na I, O I and S I, while the $H$ band is dominated by emission features produced by iron-group elements (Fe II).  The $K$ band spectra appear to contain emission features from Na I, Mg II and the first overtone of CO (see also Figure \ref{fig:K}).  We will discuss this in \S \ref{sec:CO}.   The [Fe II] lines at $\lambda$$_{rest}$= 1.257\,$\mu$m and $\lambda$$_{rest}$= 1.644\,$\mu$m are particularly prominent in the J and H bands, respectively.  These are complemented with unblended Fe II lines in the optical at expansion velocities $\sim$\,4000--6000\,$\kms$.

The broad feature positioned at $\sim$1.13\,$\mu$m can be identified with the 1.1287\,$\mu$m (3p$\ ^{3}$P-3d$\ ^{3}$D$\ ^{0}$) O I transition \citep{Meikle89}.  The Mg I (1.503\,$\mu$m) feature observed in the $H$ band is matched by strong features of Mg II at 4481\,$\AA$ in the optical region. The $K$ band region of the NIR contains the 2.207\,$\mu$m doublet of Na I.  In the optical too, Na I dominates the region around 5900\,$\AA$ from the early to very late phases. Features which appear near 1.94 and 1.98\,$\mu$m around 138\,d can probably be identified with Ca I.

\begin{figure}[t]
\centering
\includegraphics[width=0.5\textwidth, trim=4mm 0mm 0mm 0mm, clip]{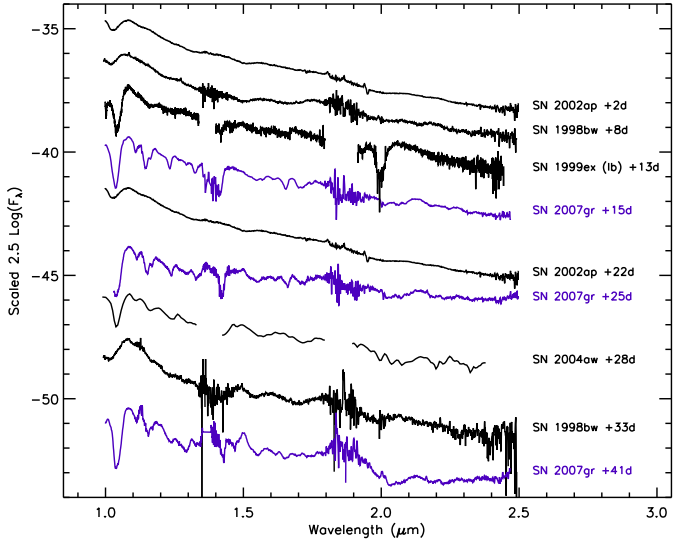}
\caption{Comparison of the near-IR spectra of SN~2007gr to SNe 2002ap (broad-lined Ic), 1998bw (broad-lined Ic), 1999ex (type Ib), and 2004aw (type Ic).  Note the presence 
of the prominent He I 1.083, and 2.0581\,$\mu$m lines in the spectrum of SN~1999ex. 
See text for references.}
\label{fig:NIR_spectra_comparison}
\end{figure}

Currently, few near-IR spectra of Ib/c SNe are found in the literature. This is particularly 
true at late phases. At earlier epochs, the situation is somewhat less dire and we compare our 
early time spectra of SN~2007gr with those of SNe~Ic 2004aw \citep{Taubenberger06}, 1998bw \citep{Patat01}, the SN~Ib \object{1999ex} \citep{Hamuy02}, and previously unpublished spectra of SN~2002ap (Ic)\footnote{These spectra were obtained at the UKIRT telescope and reduced using standard procedures.}. The near-IR spectra of all SNe mentioned above are presented in Fig. \ref{fig:NIR_spectra_comparison}.

The broad-lined spectra of SNe~2002ap and 1998bw appear almost featureless,
with the exception of a weak absorption at $\sim$1.04\,$\mu$m. This behaviour probably reflects the severe line blending due to high ejecta velocities in these objects. In contrast, the distinct narrow features of SN~2007gr are clearly visible at similar epochs.
The type Ib SN~1999ex exhibits equally strong He I absorption features at both 1.04\,$\mu$m and $\sim$2.0\,$\mu$m. This is supported by the presence of He I lines in the optical spectra \citep{Hamuy02}.  Interestingly, all SNe shown in Fig. \ref{fig:NIR_spectra_comparison} display a prominent feature near 1.04\,$\mu$m regardless of epoch, however with the exception of SN~1999ex none shows an absorption near 2.0\,$\mu$m at any time.  The 1.04\,$\mu$m absorption feature in SN~2007gr appears to be stronger than in SN~1999ex, providing some support for the contribution of C~I.

\subsection{Ejecta velocity}
\label{sec:velocity}

\begin{figure*}[t]
	\centering
	  \includegraphics[width=0.98\textwidth,angle=0.0]{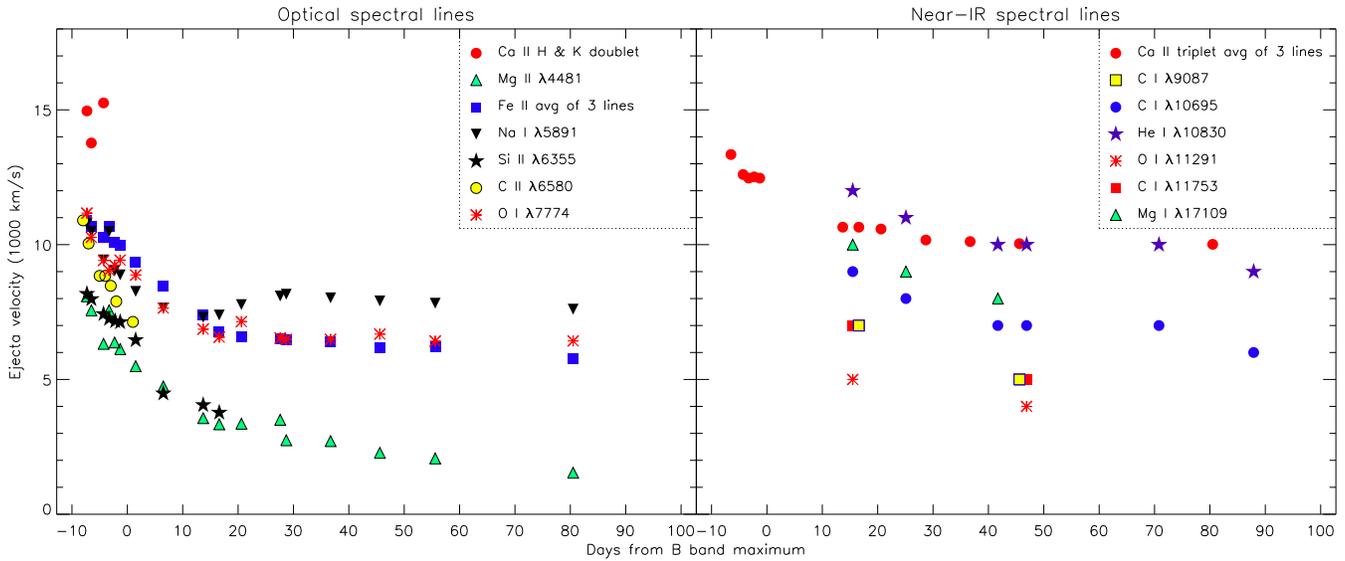}
	\caption{Evolution of the velocities from the different spectral lines observed in SN~2007gr. The errors in the measured velocities are estimated to be $\sim$\,10 \%.  The minimum error is not less than 500 \kms.  }
	\label{fig:Ejecta_velocities_07gr}
\end{figure*}

The velocities of several ejecta lines were determined by fitting a Gaussian profile to the 
absorption component of the P-Cygni feature in order to measure the position of the blue-shifted minimum.  This method provides an estimate of the expansion velocities of the layers where the individual lines predominantly form and therefore the distribution of elements within the ejecta. Only the velocities of a few spectral features have been measured since for many others line blending is a problem, even for the relatively low velocities of SN~2007gr.  The evolution of ejecta velocities is shown in Fig. \ref{fig:Ejecta_velocities_07gr}. 

The Ca II velocities of both the H \& K features as well as the near-IR triplet are significantly higher than those of Mg II (4481 $\AA$), Fe II (4924, 5018, 5170\,$\AA$), Na I (5890-5896\,$\AA$), O I (7774\,$\AA$), C I (9087, 10695 and 11753 $\AA$).  Assuming homologous expansion, the higher velocity Ca II lines imply that they form further out in the ejecta than the O I, Na I, and Fe II lines which emerge from the inner ejecta layers.  The formation of Ca II lines further out is predominantly because these lines are extremely strong and therefore become optically thick very rapidly, even in the low density, outer layers of the ejecta.  As expected, the absorption component of the P-Cygni minima gradually shift redwards with time revealing the deeper and more slow moving ejecta.

\begin{figure*}[t]
	\centering
		\includegraphics[width=0.99\textwidth,angle=0.0, trim=0mm 0mm 0mm 0mm]{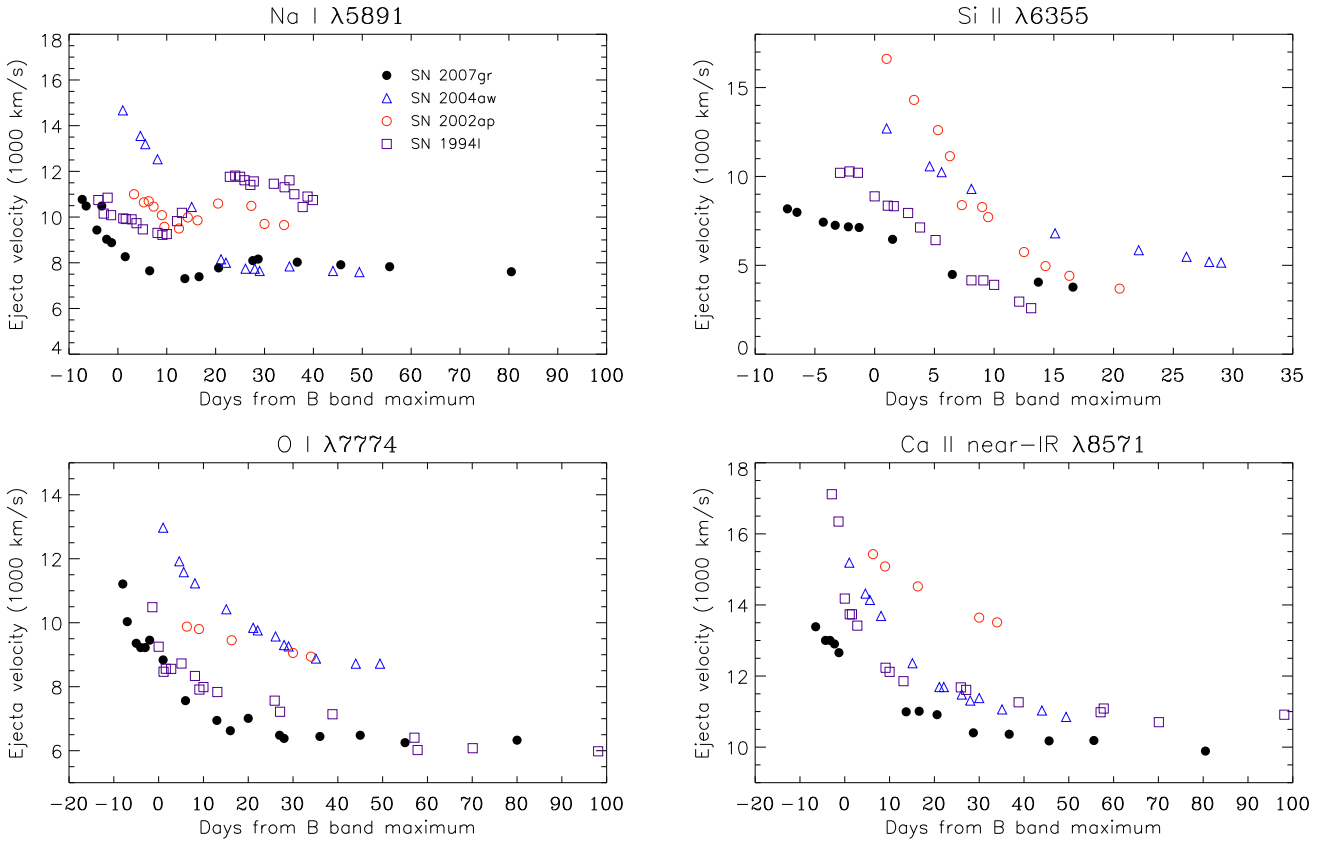}
	\caption{Comparison of the velocity evolution of the type Ic SNe~2007gr, 2004aw, 2002ap, 
                 and 1994I using different spectral lines.}
	\label{fig:Ejecta_velocities_Ic}
\end{figure*}

Comparisons of velocity widths of the Na I $\lambda$ 5891, Si II $\lambda$ 6355, O I $\lambda$ 7774 and the Ca II NIR triplet for SNe~2007gr, 2004aw, 2002ap and 1994I are shown in Fig. \ref{fig:Ejecta_velocities_Ic}.  The expansion velocity $v_{\mathrm{exp}}$ of SNe Ic can be directly measured from the minima of spectral absorption lines, of which the Si II line (6355\,$\AA$) is commonly used when present. Assuming that the blue minimum contains no contamination from other ion species, the derived $v_{\mathrm{exp}}$ values of SN~2007gr reveal comparatively low velocities which are probably related to the narrow features observed in the spectra.  Using the Si II velocity as an approximation, we estimate a photospheric velocity of $\sim$\,6,700\, \kms\, at $B$ maximum, which decreases to only $\sim$\,4,000\,\kms\, ten days later. 

 In contrast, the photospheric velocity of SN~2002ap declines rapidly and ranges from $\sim$\,17,000 to 7,000\, \kms\, in a similar time frame.  The blue-shifts of the Si II line of these two objects are comparable at two weeks from $B$ maximum, however the spectral features of SN~2002ap remain broader than those of SN~2007gr.   This suggests that the differences between SNe~2007gr and 2002ap result from the differences in their density profiles, i.e. a steeper density gradient for SN~2007gr and hence a smaller velocity range at which the lines can form.  The photospheric velocity of SN~2007gr is also similar to that of SN~1994I about one week after $B$ maximum, implying that the kinetic energy to unit mass should be also similar at this epoch. 

An interesting point to note is that the measured expansion velocities for Na I 5890, 5896\,$\AA$ about two weeks post-maximum show an increase for all Ic SNe in Fig.\ref{fig:Ejecta_velocities_Ic} with the exception of SN~2004aw, followed by a plateau.  A possible explanation for this behaviour may be the contribution of another line. A likely possibility is the contribution from He I $\lambda$\,5876\,${\AA}$ which has been previously discussed by \citet{Clocchiatti96}, although as noted previously other strong lines of He are not obvious in the optical spectra. In addition to the possibility of contamination from He I, the recombination of Na II to Na I may be an important mechanism which contributes to the strength of the Na I line in different velocity ranges.  The observed behaviour may also be explained by the departure from LTE to NLTE, again affecting the strength of the Na ID feature.

\subsection{Carbon monoxide emission}
\label{sec:CO}

Carbon monoxide emission in a supernova was first reported by \citet{Catchpole87} in \object{SN~1987A}.  Spectroscopic coverage of the first overtone ($\Delta$$\nu$=2) spanning days $\sim$\,110--574 was presented by \citet{Meikle93} while coverage of the fundamental ($\Delta$$\nu$=1) band emission spanning days $\sim$\,157--615 was presented by \citet{Bouchet93} and \citet{Wooden93}.  Since then the first overtone of CO has been observed in numerous type II SNe with detections in the spectra of \object{SN~1995ad} \citep{Spyromilio96}, \object{SN~1998S} \citep{Gerardy00,Fassia01} and \object{SNe~1998dl} $\&$ \object{1999em} \citep{Spyromilio01}.  

With the sole exception of SN~1987A, observations of the CO fundamental band (near $\sim$\,4.\,$\mu$m) have remained elusive, given the challenging nature of the observations.  However, with space-based instrumentation, the CO fundamental has also been observed in several IIP SNe: SNe~2004dj, 2005af, 2004et  \citep{Kotak05,Kotak06,Kotak09}.  Although CO emission in type II seems to be a common occurrence, detections in type Ic have been limited to SN~2000ew \citep{Gerardy02} and now SN~2007gr.

The subject of molecular emission in SNe has been extensively reviewed by \cite{Petuchowski89}, \cite{Lepp90}, \cite{Dalgarno98} and \cite{Gearhart99}, with particular reference to SN~1987A.  The main formation mechanism is the radiative association of C and O atoms whereby photons are emitted and bound CO molecules are formed.  With time, the flux of the CO emission will decrease as a result of destructive interactions with fast electrons which are produced by the $\gamma$-rays from the decay of $^{56}$Co.  Additionally, the presence of ionised He that is microscopically mixed within the carbon-oxygen layers could destroy the CO molecules via charge transfer reactions:
\begin{equation}
\hspace{2 cm} $He$^+ +~$CO$~ \rightarrow ~$C$^+ + ~$O$ ~+ ~$He$
\end{equation}
Thus molecular emission is a useful diagnostic for examining the conditions and degree of mixing in the ejecta of SNe.  

We see clear evidence for the presence of the first-overtone of CO emission from 2.29 $\mu$m to the red end of the spectrum at 2.4\,$\mu$m. This feature first appears at +\,70\,d post-$B$ maximum which is the earliest reported detection of CO in any type of SN. In part, this might be due to the lack of suitable data for the majority of SNe. However, the early appearance of CO is in line with its relatively rapid evolution of SN~2007gr compared to other SNe for which CO has been detected (e.g. SN~1987A) and is also consistent with the idea that the progenitor of SN~2007gr was a compact star.  In earlier sections we have argued against the presence of large amounts of He in the ejecta of SN~2007gr, so the fact that CO is detected at all would seem to indirectly support this idea.  Although one might invoke the possibilities that ionised He was not microscopically mixed with the underlying C/O rich layers \citep[][and references therein]{Lepp90, Gerardy00}, or that a significant quantity of He was not ionised, both possibilities seem contrived in view of the observation that strong spectral signatures due to He are not convincingly detected over a time-span of more than a year.

The CO feature grows in strength up to day +\,138 before fading.  In SN~1987A, a similar trend was observed, except that the intensity of the feature increased until $\sim$ 200\,d. This is believed to be due to the non-uniform deposition of energy at early times as a result of the greater optical depth of $\gamma$-rays in the iron core.  After sufficient expansion of the ejecta by late times, the optical depth has decreased, and greater energy deposition occurs in layers other than the iron core, i.e. the C/O layer \citep{Liu95}.  Thus the energy deposition becomes progressively uniform. 

Fig. \ref{fig:NIR_spectra} shows that by day +\,176, the CO emission has faded by a factor of $\sim$1.5 compared to day +\,138 beyond 2.35 $\mu$m. This decrease in flux may represent the enhanced energy deposition rate into the C/O layers from fast electrons.

\begin{figure}[t]
\centering
\includegraphics[width=0.50\textwidth, trim=0mm -2mm 0mm 0mm, clip]{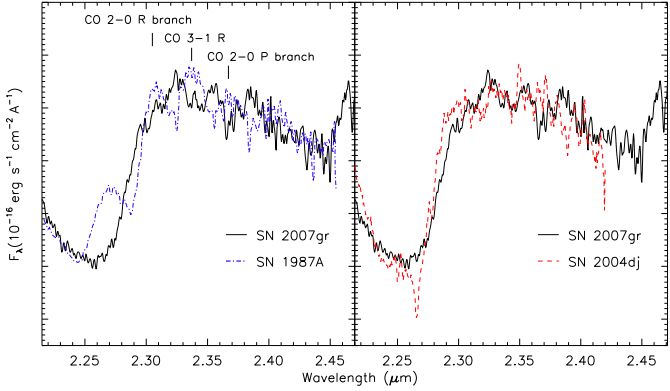}
\caption{\textbf{Left:} The CO emission profile of SN~2007gr (137.7\,d post $B$ maximum) overlaid with the 192\,d (post-explosion) spectrum of SN~1987A.  The features indicated refer to SN~1987A.
\textbf{Right:} CO profile of SN~2007gr at 137.7\,d (post-$B$-maximum) overlaid with that of
SN~2004dj at 137\,d post-explosion. }
\label{fig:K}
\end{figure}

As mentioned by \citet{Gerardy02}, the shape of the CO profile provides a lower limit to the temperature of the emitting region and so band profile modeling can be used to infer the excitation temperature.  Since observations of CO emission in type Ic are rather limited, a comparison with the molecular emission in the spectra of SN~1987A is a good alternative.  In Figure \ref{fig:K} we compare our $K$-band spectrum of SN~2007gr at day +\,137.7 to those of SN~1987A \citep{Meikle89} and the type IIP \object{SN~2004dj} \citep{Kotak05}. The spectra have been scaled so that the initial slopes of the CO emissions near 2.24 $\mu$m are relatively well matched.  In the +\,137.7 day spectrum of SN~2007gr, the CO feature decreases in flux at longer wavelengths in a similar manner to both SNe~1987A and 2004dj.  However, the first two band heads (labelled at $\sim$2.31 and 2.33 $\mu$m) of SNe~1987A and 2004dj can be clearly observed while that of SN~2007gr is not as distinct.  LTE modelling of CO has shown that the two band heads have comparable strengths at temperatures $<$\,2000\,K \citep[][]{Sharp89, Liu92}.  We thus conclude that the temperature of the CO emitting region in SN~2007gr is greater than 2000\,K.

An approximate velocity of the CO emission can be obtained by comparing our spectra of SN~2007gr with the higher resolution spectrum of SN~1987A. The slope of the 2-0 R-branch band head indicates that the expansion velocity in SN~2007gr is approximately similar to SN~1987A \citep[1800-2000\,$\kms$][]{Spyromilio88}, which was also the case for \object{SNe~1998dl} and 1999em \citep{Spyromilio01}.

CO was still detected in SN~1987A as late as 574\,d after explosion \citep{Meikle93}, whereas it had disappeared by day 355 in the type IIn SN~1998S \citep{Gerardy00}.

For SN~2007gr, we notice a fading of the feature by day +\,176, and we do not detect it in the +\,377\,d spectrum.  However, we cannot entirely rule out the presence of the CO overtone since the S/N is not sufficiently high.  As suggested by \citet{Liu98}, the disappearance of molecular emission may be due to the rapid decline in temperature as a result of the efficient cooling by CO at later times (decrease in heating rate and increasing departure from LTE).

Aside from the diagnostic constraints e.g. on microscopic mixing, that are provided by CO, the presence of molecules in the ejecta provides an additional source of cooling. This may be conducive to the formation of dust in the ejecta. One might expect the combination of molecules and low velocities observed in SN~2007gr to result in some condensation of dust grains, particularly graphite grains for the carbon-rich progenitor of SN~2007gr \citep{Valenti08a}.  However, at the epoch of our last near-IR spectrum ($\sim$1 year), we do not see any clear evidence of a rising continuum in the $K$-band. The $H$-band light curve does not show any sign of flattening but appears to follow the $^{56}$Co decay rate reasonably well (see Fig. \ref{fig:NIR_LC}).  The $H-K$ colour evolution of SN~2007gr argues against dust formation with a relatively blue H-K colour of $\sim$-1.1 mag at $\sim$ +\,400\,d.  Additionally, dust formation should result in a blueshift in the peaks of emission lines which increases with time.  We observe no such effect in the evolution of the emission features in the optical or near-IR spectra of SN~2007gr (see Fig. \ref{fig:Optical_spectra} \& \ref{fig:NIR_spectra}, respectively).

In relation to other core-collapse SNe types, IR signatures of dust condensation were found in the ejecta of the type IIP SN~1987A as early as 100\,d post-explosion \citep{Suntzeff90}. Detections were also reported at 350\,d \citep{Meikle93} and $\sim$500\,days after the outburst \citep[see e.g.][]{Bouchet93}.  Signatures of dust were evident in the type IIP SN~1999em after $\sim$\,450\,d \citep{Elmhamdi03} and also in SN~2004et starting at $\sim$300\,days \citep{sahu:06,Kotak09}.  Additionally, dust formation in the type Ib SN~1990I was reported to have commenced as early as $\sim$\,250\,d \citep{Elmhamdi04}. However, at the latest epochs under consideration here, we can conclusively state that signatures of dust formation in the ejecta of SN~2007gr are not apparent. Yet, given the rapid evolution of the SN, one might have expected signatures due to dust at epochs earlier than those for SN~1987A, and other type II-P SNe.

\subsection{Estimate of the Fe mass}

The very late-time near-IR spectrum at +\,376.7\,d (Fig. \ref{fig:NIR_spectra}) represents the only one of its kind for type Ic SNe. 

As SN~2007gr evolves into its super-nebular phase, the rapid decrease in the electron temperature and density of the ejecta implies that strong [Fe II] and [Si I] lines in the $J$ and $H$ bands should be present.  This is borne out by the observations with the prominent feature at $\sim$1.65 $\mu$m either arising from [Fe II] 1.6435 $\mu$m (a$^{4}$F-a$^{4}$D) or [Si I] 1.6454 $\mu$m, or even a blend of both.  To further add to the complexity of this scenario, the intensity ratio between the lines of these ions (i.e. [Fe II] 1.6454, 1.6068 $\mu$m and [Si I] 1.6435, 1.5994$\mu$m) are expected to be similar at high densities \citep[see][]{Oliva87}. 

However, if this feature is mainly due to [Fe II] then we should be able to detect also the 1.257 $\mu$m (a$^{6}$D-a$^{4}$D) transition, which shares the same upper level with the 1.6435 $\mu$m emission line.  In our late-time spectrum, we do not clearly detect the 1.257 $\mu$m line which should, in fact, have an intensity $\sim$1.35 times greater than the 1.6435 $\mu$m line \citep{Meikle89}.  Alternatively, if we assume [Si I] dominates the contribution to the 1.65 $\mu$m feature, we expect to observe the strongest near-IR transition of [Si I] at 1.0991 $\mu$m.  Thus the feature at $\sim$ 1.1 $\mu$m may be mainly attributed to [Si I].  The identification of [Si I] is supported by the absence of a weaker [Fe II] emission at 1.68 $\mu$m ($\sim$ 25~$\%$ the flux of [Fe II] 1.6454\,$\mu$m).  
  
 For the purpose of estimating an upper limit to the mass of $^{56}$Fe ejected in SN~2007gr , we have used the earlier spectrum at +\,175\,d where the identification of the 1.65 $\mu$m feature with [Fe II] is more secure and the S/N is much higher.

The quantity of iron in the ejecta can be derived if we know the temperature, density and emissivity of the [Fe II] emission regions \citep{Graham86}.  Therefore models which consider the ionisation structure of the SN ejecta are necessary to determine an accurate iron mass at this epoch.  \citet{Graham86} related the 1.6435 micron line flux to the iron mass (M$_{Fe}$) for the Type Ib \object{SN~1983N} according to the equation:

\begin{equation}
	\hspace{1 mm} F_{1.644 \mu m}=\frac{A_{a_{4}F_{9/2}-a_{4}D_{7/2}}hv}{4 \pi d^{2}}    
\frac{n_{a^{4}D_{7/2}}}{n_{Fe II}}     \frac{n_{Fe II}}{n_{Fe}}    \frac{M_{Fe}}{m_{Fe}}    exp(-\tau_{H})
\end{equation}

where $A$ denotes the Einstein spontaneous decay probability (0.00569\,s$^{-1}$; Nussbaumer \& Storey 1982),\textit{h}$\nu$ is the energy of the transition (i.e. a$_{4}$F$_{9/2}$-a$_{4}$D$_{7/2}$), and $d$ is the distance to the SN.  The optical depth of the line is represented by $\tau$$_{H}$, and m$_{\mathrm{Fe}}$ denotes the mass of an iron atom.  \citet{Graham86} estimated that the ejecta of SN~1983N contained $\sim$0.3 $M_{\sun}$ of iron.

To obtain an estimate for SN~2007gr, we have adopted similar physical parameters.  Assuming LTE conditions, the forbidden lines are excited at temperatures $\sim$6500K, and so n$_{a^{4}D_{7/2}}$/n$_{Fe}$ $\approx$ 0.03.  This assumption is valid for our 176\,d spectrum since electron densities are typically $>$ 10$^{6}$ cm$^{-3}$ before $\sim$ 200 days \citep{Graham86}.

Since our spectrum was observed $\sim$ half a year from $B$ maximum, the ionisation fraction n$_{Fe II}$/n$_{Fe}$ can be approximated as 0.1 \citep{Meyerott80,Axelrod80}.  Taking into account the distance to SN~2007gr and a negligible extinction for the 1.6435 $\mu$m line in the H band (see Table \ref{07gr_parameters}), the integrated line flux of $\sim$6 x10$^{-15}$ erg cm$^{-2}$s$^{-1}$ implies that 0.14 $\pm$ 0.08 $M_{\sun}$ of Fe was synthesised in SN~2007gr.  Since the quantity of $^{56}$Ni derived from the bolometric light curve is 0.076 $\pm$ 0.010 M$_{\sun}$, the iron mass is a plausible upper limit.  The error in the estimate of Fe accounts for the uncertainties in the distance to SN~2007gr, the extinction and also on the atomic data.

\section{Summary}

We have presented an extensive set of optical and near-IR observations of the type Ic SN~2007gr.
In order to put SN~2007gr into context, we compared our data with a sample of relatively well-monitored type Ic SNe that span a range in observed properties. The light curves of SN~2007gr best resemble those of the broad-lined (`hypernova') type Ic SN~2002ap at early times, peaking with a $B$ band magnitude of $M_{B}=-16.8$.  However, the peak absolute optical magnitudes lie towards the fainter end of the Ic SNe that we considered here.

The narrow spectral features of SN~2007gr compared to typical SNe Ib/c, are indicative of low expansion velocities, and facilitate line-identification which is usually problematic in type Ic SNe due to severe line-blending. The similarity with SN~2002ap does not extend to the spectroscopic observations, with SN~1994I 
generally providing a better match. 

Our nebular phase spectra reveal a systematic blue-shift in the profiles of forbidden [O I] and Mg I], while those of the permitted lines of the same elements are centred at their rest wavelengths.  This implies that the forbidden and permitted lines appear in different regions of the ejecta, as expected. The systematic blue-shift of the forbidden lines is mostly likely due to the absorption of photons emitted on the far side of the SN by the intervening,
higher opacity material.

We found low Mg I]/[O I] ratios for SN~2007gr during the early nebular phase supporting the previous suggestion that the progenitor was a carbon-rich star \citep{Valenti08a}.  This finding, taken together with the comparisons with SN~2002ap, and the relatively rapid spectral evolution e.g. the early appearance of CO, suggest a highly-evolved, compact progenitor.  From modelling of the quasi-bolometric light curve, we estimate that a modest 2-3.5\,$M_\odot$ of material was ejected by SN~2007gr at a kinetic energy of 1-4 foe, and 0.076 $\pm$ 0.020\,$M_\odot$ of $^{56}$Ni was synthesised in the explosion -- with the latter value being very similar to that found for SN~2002ap.

We clearly detected the first-overtone of the CO molecule and were able to track its evolution. It first appeared about day 71 -- the earliest recorded appearance of this molecule and persisted for at least 100\,d.

The proximity of SN~2007gr, one of the nearest SNe Ic in recent years, has provided us with an ideal target for intensive follow-up.  Indeed, SN~2007gr is the first type Ic SN for which near-IR spectra have been obtained at all phases. We suggest that SN~2007gr is an ideal template for future SNe Ib/c studies.

\section{Acknowledgements}
We would like to thank all the staff from the Asiago Ekar Telescope, Campo Imperatore Telescope, Calar Alto Observatory, Gemini North Telescope, Nordic Optical Telescope (operated on the island of La Palma jointly by Denmark, Finland, Iceland, Norway, and Sweden), Osservatorio di Teramo, Sternberg Astronomical Institute Telescope, Telescopio Nazionale Galileo, UKIRT, and the Wendelstein Telescope. The Gemini data reported here were obtained via programmes GN-2007B-DD-3 
and GN-2008B-Q-58.

We are grateful to the staff at the NOT:  Amanda Djupvik, Thierry Morel, Jarkko Niemela, Tapio Pursimo, Auni Somero, John Telting and Helena Uthas, for observing SN~2007gr.  We would also like to thank the staff at Gemini North: Thomas Dall, Tom Geballe, Silas Laycock, Atsuko Nitta, Kathy Roth, Ricardo Schiavon, Chad Trujillo and Kevin Volk, and also to the observers at the TNG: Avet Harutyunyan, Lorenzi Vania, and the Wendelstein observatory: Remus Bergemann, Florian Lang and Johannes Koppenhoefer.

J. Fynbo acknowledges the Dark Cosmology Centre which is supported by the DNRF; R.K. and S.J.S acknowledge financial support from STFC.

This paper has made use of data obtained from the Isaac Newton Group Archive which is maintained as part of the CASU Astronomical Data Centre at the Institute of Astronomy, Cambridge.
This research has made use of the NASA/IPAC Extragalactic Database (NED) which is operated by the Jet Propulsion Laboratory, California Institute of Technology, under contract with the National Aeronautics and Space Administration.
This publication has made use of data products from the Two Micron All Sky Survey, which is a joint project of the University of Massachusetts and the Infrared Processing and Analysis Center/California Institute of Technology, funded by the National Aeronautics and Space Administration and the National Science Foundation.

\begin{appendix}
\section{Photometry and Spectroscopy of SN~2007gr}

\addtocounter{table}{0}
\small
\begin{table*}[t]
\begin{center}
\caption{$U$, $B$, $V$, $R$ and $I$ magnitudes of the sequence stars in the field of NGC~1058.  The errors reported in parentheses are the rms of the measurements of the different photometric nights.   }
\begin{footnotesize}
\begin{tabular}{lcccccc}
\hline\hline
Star          & $U$ 	& $B$ 		& $V$ 		& $R$ 		& $I$\\
\hline
1	&18.531 (0.020)	&17.232 (0.009)	&16.042 (0.010)	&15.277	(0.010)	&14.655 (0.015)\\
2	&15.039	(0.010)	&15.052	(0.005)	&14.514	(0.010)	&14.184	(0.010)	&13.876 (0.012)\\
3	&16.904	(0.020)	&16.400	(0.011)	&15.544	(0.013)	&15.047	(0.011)	&14.614 (0.009)\\
4	&17.242	(0.020)	&17.230	(0.009) 	&16.578	(0.010)	&16.200	(0.008)	&15.832 (0.010)\\
5	&18.158	(0.030)	&17.684	(0.014)	&16.857	(0.014)	&16.384	(0.015)	&15.961 (0.008)\\
6	&17.920	(0.050)	&16.913	(0.018)	&15.804	(0.020)	&15.111	(0.013)	&14.476 (0.012)\\
7	&16.492	(0.030)	&15.544	(0.007)	&14.431	(0.007)	&13.835	(0.005) 	&13.300 (0.010)\\
8	&15.504	(0.012)	&15.431	(0.007)	&14.799	(0.008)	&15.125	(0.010)	&14.770 (0.008)\\
9	&14.186	(0.010)	&14.247	(0.010)	&13.675	(0.005)	&13.339	(0.009)	&13.010 (0.013)\\
10	&13.916	(0.010)	&13.876	(0.006)	&13.234	(0.010)	&12.858	(0.010)	&12.457 (0.011)\\
\hline
\end{tabular}
\\[1.6ex]
\end{footnotesize}
\label{local_stars_mags}
\end{center} 
\end{table*}

\addtocounter{table}{1}
\longtab{2}{
\small
\begin{longtable}{lllcccccc}
\caption{\label{Optical_photometry} Optical Photometry of SN~2007gr $^{a}$}\\
\hline\hline
U.T. Date & J.D. & Phase $^{b}$ & U &B &V &R &I &Instrument		\\
   		& 2,400,000+ & (days) & & & & &  & \\
\hline
\endfirsthead
\caption{continued.}\\
\hline\hline
U.T. Date & J.D. & Phase $^{b}$ & U &B &V &R &I &Instrument		\\
   		& 2,400,000+ & (days) & & & & &  & \\
\hline
\endhead
\hline
\endfoot
18/08/07 & 54330.57	& $-6.4$ 	& $13.549\pm 0.023$	& $13.987\pm 0.023$	& $13.695\pm 0.024$	& $13.519\pm 0.024$	& $13.430\pm 0.024$	&1 \\
18/08/07 & 54330.70 	& $-6.3$ 	& $13.602\pm 0.025$	& $13.821\pm 0.057$	& $13.588\pm 0.037$	& $13.535\pm 0.041$	& $13.510\pm 0.057$	&2 \\
20/08/07 &54332.68 	& $-4.3$ 	& $13.326\pm 0.024$	& $13.672\pm 0.020$	& $13.185\pm 0.030$	& $13.214\pm 0.027$	& $13.063\pm 0.029$	&3 \\
20/08/07 &54332.70 	& $-4.3$ 	& $13.348\pm 0.023$	& $13.653\pm 0.023$	& $13.214\pm 0.024$	& $13.278\pm 0.041$	& $13.083\pm 0.057$	&2 \\
21/08/07 &54333.73 	& $-3.3$ 	& $13.258\pm 0.023$	& $13.549\pm 0.022$	& $13.086\pm 0.023$	& $13.045\pm 0.025$	& $\ldots$ 			&3 \\
22/08/07 &54334.63 	& $-2.4$ 	& $13.291\pm 0.025$	& $13.481\pm 0.025$	& $13.046\pm 0.024$	& $12.887\pm 0.023$	& $12.858\pm 0.024$	&4 \\
22/08/07 &54334.70 	& $-2.3$ 	& $\ldots$				& $13.509\pm 0.023$	& $13.000\pm 0.033$	& $12.969\pm 0.026$	& $\ldots$				&3 \\
23/08/07 &54335.74 	& $-1.3$	& $13.345\pm 0.025$	& $13.512\pm 0.037$	& $12.981\pm 0.028$	& $12.919\pm 0.054$	& $12.839\pm 0.069$	&3 \\
24/08/07 &54336.51 	& $-0.5$ 	& $13.340\pm 0.026$	& $13.511\pm 0.021$	& $12.947\pm 0.020$	& $12.837\pm 0.020$	& $12.731\pm 0.020$	&4 \\
24/08/07 &54336.60 	& $-0.4$ 	& $\ldots$				& $13.438\pm 0.025$	& $12.960\pm 0.025$	& $12.834\pm 0.026$	& $12.674\pm 0.026$	&5 \\
26/08/07 &54338.53 	& $+1.5$	& $\ldots$				& $13.495\pm 0.025$	& $12.882\pm 0.023$	& $12.732\pm 0.024$	& $12.717\pm 0.023$	&4 \\
26/08/07 &54338.60 	& $+1.6$	& $\ldots$				& $13.468\pm 0.026$	& $12.904\pm 0.026$	& $12.770\pm 0.032$	& $12.645\pm 0.026$	&5 \\
26/08/07 &54339.43 	& $+2.4$	& $\ldots$				& $13.580\pm 0.030$	& $12.916\pm 0.027$	& $12.810\pm 0.045$	& $\ldots$ 			&6 \\
27/08/07 &54339.60 	& $+2.6$	& $\ldots$				& $13.531\pm 0.021$	& $12.908\pm 0.020$	& $12.741\pm 0.032$	& $12.620\pm 0.022$	&5 \\
29/08/07 &54342.37 	& $+5.4$	& $14.182\pm 0.030$	& $13.730\pm 0.022$	& $13.049\pm 0.030$	& $12.787\pm 0.024$	& $12.611\pm 0.040$	&6 \\
30/08/07 &54342.52 	& $+5.5$	& $\ldots$ 			& $\dots$				& $13.070\pm 0.100$	& $12.780\pm 0.060$	& $12.608\pm 0.030$	&4 \\
31/08/07 &54344.28 	& $+7.3$	& $\ldots$				& $13.939\pm 0.100$	& $13.163\pm 0.100$	& $12.832\pm 0.100$	& $12.650\pm 0.100$	&6 \\
03/09/07 &54347.27 	& $+10.3$& $14.890\pm 0.150$	& $14.200\pm 0.150$	& $13.265\pm 0.150$	& $12.897\pm 0.150$	& $12.736\pm 0.150$	&6 \\
06/09/07 &54350.49 	& $+13.5$& $\ldots$				& $14.580\pm 0.022$	& $13.579\pm 0.023$	& $13.097\pm 0.023$	& $12.766\pm 0.022$	&1 \\
07/09/07 &54350.67 	& $+13.7$& $15.280\pm 0.026$	& $14.443\pm 0.044$	& $13.516\pm 0.028$	& $13.154\pm 0.055$	& $12.827\pm 0.031$	&3 \\
10/09/07 &54353.67 	& $+16.7$& $15.613\pm 0.030$	& $14.970\pm 0.023$	& $13.881\pm 0.026$	& $13.348\pm 0.030$	& $12.932\pm 0.033$	&3 \\
11/09/07 &54355.34 	& $+18.3$& $15.918\pm 0.026$	& $15.122\pm 0.100$	& $13.974\pm 0.100$	& $13.551\pm 0.050$	& $13.008\pm 0.050$	&6 \\
13/09/07 &54356.60 	& $+19.6$& $\ldots$				& $15.248\pm 0.028$	& $14.088\pm 0.025$	& $13.528\pm 0.023$	& $13.111\pm 0.025$	&5 \\
14/09/07 &54357.52 	& $+20.5$& $\ldots$				& $15.261\pm 0.026$	& $14.078\pm 0.021$	& $13.535\pm 0.022$	& $13.118\pm 0.023$	&7 \\
14/09/07 &54357.55 	& $+20.6$& $\ldots$				& $15.250\pm 0.024$	& $14.098\pm 0.023$	& $13.589\pm 0.025$	& $13.219\pm 0.023$	&1 \\
14/09/07 &54357.60 	& $+20.6$& $\ldots$				& $15.393\pm 0.028$	& $14.156\pm 0.020$	& $13.608\pm 0.026$	& $13.207\pm 0.026$	&5 \\
15/09/07 &54358.53 	& $+21.5$& $16.198\pm 0.054$	& $15.365\pm 0.032$	& $14.184\pm 0.023$	& $13.592\pm 0.024$	& $13.185\pm 0.022$	&4 \\
15/09/07 &54358.57 	& $+21.6$& $16.125\pm 0.024$	& $15.423\pm 0.024$	& $14.202\pm 0.023$	& $13.624\pm 0.023$	& $13.231\pm 0.024$	&3 \\
16/09/07 &54359.55 	& $+22.6$& $\ldots$				& $\ldots$				& $\ldots$				& $13.710\pm 0.020$	& $13.281\pm 0.021$	&5 \\
16/09/07 &54360.45 	& $+23.5$& $16.275\pm 0.041$	& $15.427\pm 0.032$	& $14.309\pm 0.027$	& $13.743\pm 0.028$	& $13.252\pm 0.024$	&4 \\
17/09/07 &54360.64	 	& $+23.6$& $\ldots$				& $15.531\pm 0.015$	& $14.325\pm 0.017$	& $13.752\pm 0.023$	& $13.299\pm 0.006$	&5 \\
18/09/07 &54362.44 	& $+25.4$& $16.413\pm 0.045$	& $15.645\pm 0.030$	& $14.380\pm 0.025$	& $13.856\pm 0.023$	& $13.405\pm 0.027$	&4 \\
20/09/07 &54363.60 	& $+26.6$& $\ldots$				& $15.654\pm 0.021$	& $14.530\pm 0.015$	& $13.959\pm 0.020$	& $13.456\pm 0.021$	&5 \\
20/09/07 &54364.45 	& $+27.5$& $16.456\pm 0.051$	& $15.737\pm 0.032$	& $14.545\pm 0.030$	& $13.969\pm 0.024$	& $13.468\pm 0.023$	&4 \\
21/09/07 &54364.50 	& $+27.5$& $\ldots$				& $15.710\pm 0.021$	& $14.567\pm 0.014$	& $14.004\pm 0.021$	& $13.480\pm 0.028$	&5 \\
22/09/07 &54365.64 	& $+28.6$& $16.258\pm 0.023$	& $15.818\pm 0.023$	& $14.567\pm 0.023$	& $14.055\pm 0.023$	& $13.511\pm 0.023$	&3 \\
25/09/07 &54368.56 	& $+31.6$& $16.346\pm 0.080$	& $15.791\pm 0.035$	& $14.710\pm 0.030$	& $14.138\pm 0.028$	& $13.594\pm 0.025$	&4 \\
26/09/07 &54370.43 	& $+33.4$& $16.467\pm 0.100$	& $15.793\pm 0.031$	& $14.788\pm 0.027$	& $14.200\pm 0.025$	& $13.654\pm 0.021$	&4 \\
28/09/07 &54371.57 	& $+34.6$& $16.417\pm 0.024$	& $15.810\pm 0.024$	& $14.859\pm 0.024$	& $14.324\pm 0.030$	& $13.686\pm 0.033$	&3 \\
30/09/07 &54373.60 	& $+36.6$& $16.337\pm 0.032$	& $15.916\pm 0.025$	& $14.875\pm 0.021$	& $14.266\pm 0.023$	& $13.787\pm 0.025$	&3 \\
02/10/07 &54375.53 	& $+38.5$& $\ldots$				& $15.910\pm 0.025$	& $14.906\pm 0.021$	& $14.367\pm 0.020$	& $13.818\pm 0.032$	&5 \\
08/10/07 &54381.51 	& $+44.5$& $\ldots$				& $15.952\pm 0.026$	& $15.001\pm 0.028$	& $14.499\pm 0.037$	& $13.888\pm 0.031$	&5 \\
09/10/07 &54382.60 	& $+45.6$& $\ldots$				& $15.920\pm 0.024$	& $15.055\pm 0.026$	& $14.493\pm 0.026$	& $13.937\pm 0.023$	&1 \\
09/10/07 &54382.56 	& $+45.6$& $\ldots$				& $15.972\pm 0.038$	& $15.010\pm 0.018$	& $14.565\pm 0.024$	& $13.951\pm 0.023$	&5 \\
10/10/07 &54383.57 	& $+46.6$& $\ldots$				& $16.001\pm 0.025$	& $15.021\pm 0.025$	& $14.544\pm 0.021$	& $13.922\pm 0.022$	&5 \\
11/10/07 &54384.57 	& $+47.6$& $\ldots$				& $15.976\pm 0.023$	& $15.054\pm 0.018$	& $14.562\pm 0.020$	& $13.929\pm 0.029$	&5 \\
12/10/07 &54385.63 	& $+48.6$& $\ldots$				& $16.025\pm 0.026$	& $15.053\pm 0.021$	& $14.570\pm 0.020$	& $13.928\pm 0.020$	&5 \\
16/10/07 &54389.55 	& $+52.6$& $\ldots$				& $16.021\pm 0.022$	& $15.163\pm 0.018$	& $14.621\pm 0.038$	& $13.984\pm 0.028$	&5 \\
16/10/07 &54390.40 	& $+53.4$& $16.542\pm 0.040$	& $16.052\pm 0.030$	& $15.120\pm 0.025$	& $14.660\pm 0.028$	& $14.045\pm 0.020$	&4 \\
19/10/07 &54392.60 	& $+55.6$& $\ldots$				& $15.984\pm 0.028$	& $15.162\pm 0.025$	& $14.680\pm 0.022$	& $14.060\pm 0.020$	&1 \\
21/10/07 &54394.52 	& $+57.5$& $16.571\pm 0.038$	& $16.058\pm 0.032$	& $15.201\pm 0.026$	& $14.743\pm 0.023$	& $14.127\pm 0.025$	&4 \\
23/10/07 &54396.52 	& $+59.5$& $\ldots$				& $16.073\pm 0.023$	& $15.235\pm 0.024$	& $14.772\pm 0.021$	& $14.133\pm 0.022$	&4 \\
24/10/07 &54398.48 	& $+61.5$& $16.591\pm 0.037$	& $16.142\pm 0.029$	& $15.267\pm 0.022$	& $14.853\pm 0.024$	& $14.168\pm 0.025$	&4 \\
29/10/07 &54402.58 	& $+65.6$& $16.690\pm 0.043$	& $16.168\pm 0.031$	& $15.319\pm 0.025$	& $14.906\pm 0.024$	& $14.216\pm 0.023$	&4 \\
30/10/07 &54403.58 	& $+66.6$& $\ldots$				& $16.175\pm 0.029$	& $15.352\pm 0.021$	& $14.905\pm 0.020$	& $14.238\pm 0.020$	&5 \\
02/11/07 &54406.70 	& $+69.7$& $16.662\pm 0.042$	& $16.209\pm 0.026$	& $15.379\pm 0.024$	& $14.973\pm 0.025$	& $14.241\pm 0.021$	&4 \\
06/11/07 &54410.70 	& $+73.7$& $16.740\pm 0.036$	& $16.295\pm 0.025$	& $15.458\pm 0.021$	& $15.046\pm 0.023$	& $14.351\pm 0.023$	&4 \\
13/11/07 &54417.58 	& $+80.6$& $\ldots$				& $16.295\pm 0.031$	& $15.600\pm 0.022$	& $15.133\pm 0.023$	& $14.518\pm 0.025$	&1 \\
14/11/07 &54418.56 	& $+81.6$& $16.811\pm 0.051$	& $16.378\pm 0.041$	& $15.579\pm 0.030$	& $15.177\pm 0.027$	& $14.480\pm 0.032$	&4 \\
21/11/07 &54425.63 	& $+88.6$& $\ldots$				& $16.472\pm 0.020$	& $\ldots$				& $15.215\pm 0.020$	& $\ldots$				&5 \\
27/11/07 &54432.44 	& $+95.4$& $17.027\pm 0.045$ 	& $16.581\pm 0.037$	& $15.824\pm 0.022$	& $15.388\pm 0.026$	& $14.724\pm 0.028$	&4 \\
04/12/07 &54439.51 	&$+102.5$&$\ldots$				& $16.607\pm 0.040$	& $15.923\pm 0.031$	& $15.427\pm 0.025$	& $14.825\pm 0.023$	&1 \\
05/12/07 &54440.45 	&$+103.5$&$\ldots$				& $16.665\pm 0.035$	& $15.938\pm 0.025$	& $15.442\pm 0.023$	& $14.922\pm 0.024$	&1 \\
06/12/07 &54440.53 	&$+103.5$&$17.094\pm 0.040$	& $16.657\pm 0.032$	& $15.935\pm 0.023$	& $15.506\pm 0.021$	& $14.859\pm 0.023$	&4 \\
10/12/07 &54444.51 	&$+107.5$&$\ldots$				& $16.745\pm 0.028$	& $16.039\pm 0.026$	& $15.519\pm 0.021$	& $15.000\pm 0.021$	&5 \\
10/12/07 &54444.54 	&$+107.5$&$17.203\pm 0.060$	& $16.761\pm 0.032$	& $16.028\pm 0.022$	& $15.570\pm 0.022$	& $15.004\pm 0.025$	&4 \\
16/12/07 &54450.62 	&$+113.6$&$17.188\pm 0.070$	& $16.830\pm 0.040$	& $16.122\pm 0.030$	& $15.626\pm 0.027$	& $15.058\pm 0.028$	&2 \\
25/12/07 &54460.49 	&$+123.5$&$\ldots$				& $16.970\pm 0.023$	& $16.326\pm 0.023$	& $15.709\pm 0.026$	& $15.206\pm 0.026$	&5 \\
28/12/07 &54463.31 	&$+126.3$&$\ldots$				& $17.030\pm 0.029$	& $\ldots$				& $15.781\pm 0.026$	& $15.278\pm 0.026$	&5 \\
28/12/07 &54463.39 	&$+126.4$&$17.327\pm 0.060$	& $17.022\pm 0.047$	& $16.346\pm 0.025$	& $15.781\pm 0.027$	& $15.226\pm 0.024$	&1 \\
03/01/08 &54468.50 	&$+131.5$&$\ldots$				& $17.077\pm 0.031$	& $16.476\pm 0.024$	& $15.803\pm 0.021$	& $15.343\pm 0.021$	&5 \\
06/01/08 &54472.40 	&$+135.4$&$\ldots$				& $17.180\pm 0.045$	& $16.495\pm 0.031$	& $15.935\pm 0.025$	& $15.442\pm 0.025$	&1 \\
09/01/08 &54474.51 	&$+137.5$&$\ldots$				& $17.238\pm 0.037$	& $16.568\pm 0.024$	& $15.922\pm 0.022$	& $15.419\pm 0.020$	&5 \\
13/01/08 &54479.40 	&$+142.4$&$17.808\pm 0.055$	& $17.239\pm 0.040$	& $16.725\pm 0.037$	& $15.959\pm 0.030$	& $15.538\pm 0.027$	&3 \\
26/01/08 &54492.35 	&$+155.4$&$17.871\pm 0.131$	& $17.392\pm 0.046$	& $16.853\pm 0.028$	& $16.150\pm 0.025$	& $15.810\pm 0.026$	&8 \\
28/01/08 &54494.36 	&$+157.4$&$\ldots$				& $17.477\pm 0.047$	& $16.967\pm 0.024$	& $16.161\pm 0.020$	& $15.829\pm 0.020$	&5 \\
10/02/08 &54507.45 	&$+170.5$&$\ldots$				& $17.671\pm 0.040$	& $\ldots$				& $16.371\pm 0.027$	& $16.168\pm 0.025$	&5 \\
29/08/08 &54707.97 	&$+371.0$&$\ldots$				& $\ldots$ 			& $20.300\pm 0.200$ 	& $19.770\pm 0.200$	& $19.620\pm 0.200$	&9 \\
02/09/08 &54711.62 	&$+374.6$&$\ldots$				& $\ldots$ 			& $20.400\pm 0.500$ 	& $19.800\pm 0.500$	& $19.540\pm 0.400$	&8 \\
30/09/08 &54740.49 	&$+403.5$&$\ldots$				& $\ldots$ 			& $20.696\pm 0.500$ 	& $20.205\pm 0.400$	& $19.861\pm 0.600$	&8 \\
\hline
\end{longtable}
\begin{flushleft}
\textit{a} \hspace{2 mm} The S-corrected optical photometry of SN~2007gr.  The reported errors (in parentheses) are a result of the uncertainty due to the PSF fitting of the SN magnitude and to the background contamination (computed using the artificial-star experiment).
\newline
\hspace{2 cm} \textit{b} \hspace{2 mm} Relative to the epoch of $B$ maximum JD = 2454337.0 $\pm$ 1.0.  
\newline

\hspace{0.5 cm}1= Ekar: 1.82\,m Copernico Telescope, INAF - Osservatorio di Asiago, Mt. Ekar, Asiago (Italy) + AFOSC. 

\hspace{0.5 cm}2= TNG: 3.5\,m Telescopio Nazionale Galileo, La Palma, Canary Isl. (Spain) + DOLORES. 

\hspace{0.5 cm}3= NOT: 2.5\,m Nordic Optical Telescope, La Palma, Canary Isl. (Spain) + ALFOSC.

\hspace{0.5 cm}4= LT: 2.0\,m Liverpool Telescope, La Palma, Canary Isl. (Spain) + RATCAM.

\hspace{0.5 cm}5= Wendelstein: 0.8\,m Telescope at the Wendelstein Observatory, (Munich, Germany) + MONICA.

\hspace{0.5 cm}6= S70: 0.70\,m Sternberg Astronomical Institute Telescope, Moscow (Russia).

\hspace{0.5 cm}7= G. D. Cassini: 1.52\,m Telescope at the Loiano Observatory (BOL., Italy) + BFOSC.

\hspace{0.5 cm}8= CA: 2.2\,m  Telescope at the Calar Alto Observatory (Spain) + CAFOS.

\hspace{0.5 cm}9= Gemini North: 8.1\,m Telescope, Mauna Kea (Hawaii) + GMOS-N.

\label{Optical_photometry}
\end{flushleft}
}

\addtocounter{table}{0}
\begin{table*}
\caption{Near-Infrared Photometry of SN~2007gr $\rlap{$^{a}$}$}
\begin{footnotesize}
\begin{center}
\begin{tabular}{ccccccc}
\hline\hline
U.T. Date & J.D. & Phase $^{b}$ & J &H &K& Instrument  \\
  & 2,400,000+ & (days) & & & & \\
\hline
22/08/07 	&54335.74 	& $-1.3$ 		& $12.570\pm 0.100$	&$\ldots$					&$\ldots$				&1	\\
02/09/07 	&54347.50 	& $+10.5$		& $12.471\pm 0.030$	& $12.306\pm 0.030$		& $12.077\pm 0.030$	&2 	\\
05/09/07 	&54349.50 	& $+12.5$ 	& $12.518\pm 0.020$	& $12.307\pm 0.020$		& $12.154\pm 0.030$	&3	 \\
18/09/07 	&54361.62 	& $+24.6$ 	& $\ldots$				& $12.760\pm 0.040$		& $\ldots$				&4	 \\		
19/09/07 	&54363.60 	& $+26.6$ 	& $13.069\pm 0.020$	& $12.723\pm 0.030$		& $12.651\pm 0.030$	&5	 \\
08/10/07   &54382.60 	& $+45.6$ 	& $13.733\pm 0.040$	& $13.434\pm 0.040$		& $13.377\pm 0.040$	&1	 \\
10/10/07   &54383.39 	& $+46.4$ 	& $\ldots$				& $13.440\pm 0.050$		& $\ldots$				&4	 \\				
14/10/07 	&54388.60 	& $+51.6$ 	& $13.913\pm 0.040$	& $13.507\pm 0.040$		& $13.681\pm 0.040$	&1	 \\
17/10/07 	&54391.60 	& $+54.6$ 	& $14.020\pm 0.040$	& $13.657\pm 0.040$		& $13.674\pm 0.040$	&1	 \\
24/10/07 	&54398.60 	& $+61.6$ 	& $14.239\pm 0.040$	& $13.807\pm 0.040$		& $13.877\pm 0.040$	&1	 \\
28/10/07 	&54402.60 	& $+65.6$ 	& $14.411\pm 0.040$	& $13.911\pm 0.040$		& $13.957\pm 0.040$	&1	 \\
03/11/07 &54407.32 	& $+70.3$ 	& $\ldots$				& $14.030\pm 0.030$		& $\ldots$				&4	 \\		
20/11/07 &54424.44 	& $+87.4$ 	& $\ldots$				& $14.490\pm 0.030$		& $\ldots$				&4	 \\		
05/12/07 &54439.70 	& $+102.7$	& $15.396\pm 0.040$	& $14.795\pm 0.030$		& $15.309\pm 0.030$	&3	 \\
16/12/07 &54450.70 	& $+113.7$	& $15.860\pm 0.040$	& $15.120\pm 0.030$		& $15.415\pm 0.030$	&3	 \\
20/12/07 &54454.35 	& $+117.4$ 	& $\ldots$				& $15.240\pm 0.040$		& $\ldots$				&4	 \\
09/01/08 &54474.20 	& $+137.2$ 	& $\ldots$				& $15.590\pm 0.040$		& $\ldots$				&4	 \\
09/01/08 &54474.72 	& $+137.7$ 	& $16.306\pm 0.100$	& $15.552\pm 0.100$		& $15.603\pm 0.100$	&5	 \\
16/02/08 &54512.22 	& $+175.2$ 	& $\ldots$				& $16.250\pm 0.040$		& $\ldots$				&4	 \\
20/02/08 &54516.23 	& $+179.2$ 	& $\ldots$				& $16.320\pm 0.040$		& $\ldots$				&4	 \\
05/09/08 &54715.13 	& $+378.1$ 	& $19.214\pm 0.060$	& $\ldots$					& $\ldots$				&4	\\
06/09/08 &54716.14 	& $+379.1$ 	&  $\ldots$			& $18.196\pm 0.029$		& $\ldots$				&4	\\
08/09/08 &54718.04 	& $+381.0$ 	&  $\ldots$			& $\ldots$					& $19.280\pm 0.200$	&4	\\
\hline
\end{tabular}
\end{center}
\textit{a} \hspace{2 mm} The reported errors (in parentheses) are a result of the uncertainty due to the PSF fitting of the SN magnitude and to the background contamination (computed using the artificial-star experiment).
\newline
\textit{b} \hspace{2 mm} Relative to the epoch of $B$ maximum (JD = 2454337.0 $\pm$ 1.0).
\newline

  	 \hspace{0.5 cm} 1= CI: 1.1\,m  Telescope AZT-24, Campo Imperatore (Italy) + SWIRCAM.
	 
	 \hspace{0.5 cm} 2= NOT: 2.5\,m Nordic Optical Telescope, La Palma, Canary Isl. (Spain) + NOTCAM.
   
	 \hspace{0.5 cm} 3= TNG: 3.5\,m Telescopio Nazionale Galileo, La Palma, Canary Isl. (Spain) + NICS. 

  	 \hspace{0.5 cm} 4= Gemini North: 8.1\,m Telescope, Mauna Kea (Hawaii) + NIRI.
   
   	 \hspace{0.5 cm} 5= UKIRT: 3.8\,m Infrared Telescope, Mauna Kea (Hawaii) + WFCAM.
\end{footnotesize}
\label{NIR_photometry}
\end{table*}

\addtocounter{table}{0}
\begin{table*}[t]
\caption{Log of Optical Spectroscopy for SN~2007gr.} 
\label{parameters7}
\begin{footnotesize}
\begin{center}
\begin{tabular}{ccccccccc}
\hline
	U.T. Date  	&JD  			& Epoch $^{a}$  		&Instrumental        				&Resolution		&Exptime		&Range		&Reference $^{c}$    \\
			 	&(2,400,000+)    	& (days)				&configuration	      				&({$\AA$}) $^{b}$	&(s)			&({$\AA$})	&         			\\
       				&          			&               				&           						&                			&            		&             		&                      		  \\
         \hline   
2007 Aug 17 		&54329.7			&-7.3				&TNG+DOLORES+gm.LRS+LRB	&12				&900		&3200-7710	&1				\\
2007 Aug 17 		&54329.7			&-7.3				&NOT+ALFOSC+gm.4			&12				&250		&3400-9060	&2				\\
2007 Aug 18 		&54330.5			&-6.5				&Ekar+AFOSC+gm.4,gm.2		&15-25			&1200		&3490-9980	&1				\\
2007 Aug 20 		&54332.7        		&-4.3         				&NOT+ALFOSC+gm.4       		&12 				&900	   	&3240-9110	&1     			 \\
2007 Aug 20 		&54332.7        		&-4.3         				&TNG+DOLORES+gm.LRS+LRB  	&12 				&200	   	&3180-9400	&1     			 \\
2007 Aug 21		&54333.7			&-3.3				&NOT+ALFOSC+gm.4			&12				&300		&3170-9100	&1				\\
2007 Aug 22 		&54334.7			&-2.3				&NOT+ALFOSC+gm.4			&12				&200		&3190-9110	&1				\\
2007 Aug 23 		&54335.7			&-1.3				&NOT+ALFOSC+gm.4			&12				&300		&3130-9100	&1				\\
2007 Aug 26 		&54338.5			&+1.5				&Pennar +B$\&$C+300tr/mm		&25				&600		&3270-7810	&1				\\
2007 Aug 26		&54339.7			&+2.7				&NOT+ALFOSC+gm.4			&12				&600		&3190-9100	&2				\\
2007 Aug 31 		&54344.5			&+7.5				&Pennar +B$\&$C+300tr/mm		&25				&1200		&3450-7800	&1				\\
2007 Sep 07 		&54350.7			&+13.7				&NOT+ALFOSC+gm.4			&12				&600		&3200-8940	&1				\\
2007 Sep 10 		&54353.6			&+16.6				&NOT+ALFOSC+gm.4			&12				&600		&3200-9060	&1				\\
2007 Sep 14 		&54357.6			&+20.6				&Loiano+BIFOSC+gm.4,5		&15-25			&900	   	&3500-9780	&1				\\
2007 Sep 14		&54357.5			&+20.5				&Ekar+AFOSC+gm.4,2			&15				&1200		&3550-9830	&2				\\
2007 Sep 15		&54358.6			&+21.6				&NOT+ALFOSC+gm.4			&12				&900		&3070-9270	&2				\\
2007 Sep 21 		&54364.6   		&+27.6        			&Pennar +B$\&$C+300tr/mm        	&25 				&3000      	   	&3270-7830   	&2      			 \\
2007 Sep 22 		&54365.7   		&+28.7       			&NOT+ALFOSC+gm.4                	&12				&900       	   	&3180-9070 	&2      			 \\
2007 Sep 25		&54368.6			&+31.6				&NOT+ALFOSC+gm.4			&12				&900		&3200-9100	&2				\\

2007 Sep 30 		&54373.7			&+36.7				&NOT+ALFOSC+gm.4			&12				&900		&3240-9070	&1				\\
2007 Oct 09 		&54382.6       		&+45.6       			&Ekar+AFOSC+gm.4,2	        		&15-25             		&2400      	  	&3490-10050 	&2    				 \\
2007 Oct 19 		&54392.6        		&+55.6        			&Ekar+AFOSC+gm.4,2	        		&15-25             		&1800      	  	&3540-9860  	&2        			 \\ 
2007 Nov 13 		&54417.5        		&+80.5        			&Ekar+AFOSC+gm.4,2	        		&15-25             		&2700      	   	&3490-9980  	&2       			 \\
2007 Dec 05		&54440.4        		&+103.4       			&Ekar+AFOSC+gm.4,2	       		&15-25            		&2700      	   	&3490-10070 	&2       			 \\
2007 Dec 15 		&54450.5        		&+113.5       			&TNG+DOLORES+VHR+LRS	 	&5              		&1800      	  	&4660-6640 	&2        			 \\
2008 Jan 06 		&54472.3	      		&+135.3     			&Ekar+AFOSC+gm.4,2			&15-25       		&1800      	   	&3490-10180	&2        			 \\ 
2008 Jan 29		&54495.4        		&+158.4      			&CA+CAFOS+gm.b200,r200 		&12     	 		&2400x2      	&3210-10630	&2, 3       			  \\ 
2008 Sep 02		&54711.5			&+374.5				&CA+CAFOS+gm.g200			&12				&2700x3		&3670-10460	&2				\\
2008 Sep 03		&54712.5        		&+375.5      			&GN+GMOS+R150 				&12      			&5400      	   	&4000-11000	&2        			 \\ 
\hline
\end{tabular}
\end{center}
\textit{a} \hspace{2 mm} Relative to the epoch of $B$ maximum (JD = 2,454,337.0 $\pm$ 1.0).
\newline
\textit{b} \hspace{2 mm} Derived from FWHM of night-sky emission lines.
\newline
\textit{c} \hspace{3 mm}References:
\hspace{1 mm} [1] \citet{Valenti08a} 
\hspace{1 mm} [2] This paper
\hspace{1 mm} [3] \citet{Valenti09}. 
\newline

\hspace{0.5 cm}TNG = 3.5\,m Telescopio Nazionale Galileo, La Palma, Canary Isl. (Spain) + DOLORES.

\hspace{0.5 cm}Ekar = 1.82\,m Copernico Telescope, INAF -Osservatorio di Asiago, Mt Ekar, Asiago (Italy) + AFOSC.

\hspace{0.5 cm}NOT = 2.5\,m Nordic Optical Telescope, La Palma, Canary Isl. (Spain) + ALFOSC.

\hspace{0.5 cm}Pennar = 1.22\,m Telescope, Asiago Pennar, (Italy) + Boller $\&$ Chivens (B\&C) spectrograph.

\hspace{0.5 cm}Loiano = 1.52.\,m Telescope, INAF -Bologna Observatory, Bologna (Italy) + BIFOSC.

\hspace{0.5 cm}CA = 2.2\,m Telescope at Calar Alto Observatory (Spain) + CAFOS.

\hspace{0.5 cm}GN = 8.\,m Telescope at Gemini North, Hawaii + GMOS-N.

\end{footnotesize}
\label{log_optical_spectra}
\end{table*}

\addtocounter{table}{0}
\begin{table*}[t]
\caption{Log of Near-Infrared Spectroscopy of SN~2007gr.} 
\label{parameters8}
\begin{footnotesize}
\begin{center}
\begin{tabular}{lccccccc}
\hline
U.T. Date       & J.D    & Epoch  $^{a}$       & Standard star & t$_{exp}$/band SN $^{b}$ &Instrumental & Resolution	\\
&(2,400,000+)   &(days)  & (Spectral type)     &  (s)  & configuration $^{c}$		& $({\AA})^{d}$			\\
\hline
2007 Sep 09	&  54352.5	 &+15.5			& HIP 1603  (AOVs) 		& 300 					&TNG+NICS+gm.IJ,HK			&15-30\\
2007 Sep 18     &  54362.1        &+25.1        			& HD 19301  (F3V)            & 300 					&GN+NIRI+JHK				&20-30\\
2007 Oct 05      &  54378.7        &+41.7        			& HIP 1603  (AOVs)   	& 300    					&TNG+NICS+gm.IJ,HK		&15-30\\
2007 Oct 10      &  54383.9        &+46.9        			& HIP 12181 (F5V)   		& 300    					&GN+NIRI+JHK		&20-30\\
2007 Nov 03     &  54407.8        &+70.8        			& HD 17119  (F5V)            & 300    					&GN+NIRI+JHK			&20-30\\
2007 Nov 20     &  54424.9        &+87.9       			& HD 19301  (F5V)            & 300    					&GN+NIRI+JHK			&20-30\\
2007 Dec 16     &  54451.5        &+114.5       		& HIP 10559 (AOV)           & 600		 			&TNG+NICS+gm.IJ,HK				&15-30\\
2007 Dec 20     &  54454.8        &+117.8       		& HD 17119  (F5V)            & \hspace{2mm}840 $*$	&GN+NIRI+JHK						&20-30\\
2008 Jan 09     &  54474.7        &+137.7       			& HD 17119  (F5V)            & 840   					&GN+NIRI+JHK 			&20-30\\
2008 Feb 16     &  54512.7        &+175.7       			& HD 17119  (F5V)            & 1120   					&GN+NIRI+JHK			&20-30\\
2008 Sep 04     &  54713.6        &+376.6       			& HIP 12719  (B3V)            & 1800  					&GN+NIRI+JHK			&20-30\\
\hline
\end{tabular}
\end{center}

\textit{a} \hspace{2 mm} Epoch with respect to the estimated $B$-band maximum (JD 2,454,337.0 $\pm$ 1.0).

\textit{b} \hspace{2 mm} The quantity t$_{exp}$/band is the on-source integration time for each band.

\hspace{4.5 mm} \textit{*} \hspace{1 mm}  1050\,s exposure for the $J$ band.

\textit{c} \hspace{2 mm} TNG= 3.5\,m Telescopio Nazionale Galileo, La Palma, Canary Isl. (Spain), \hspace{0.2 cm}GN= 
8.1\,m Telescope at Gemini North, Hawaii.  

\textit{d} \hspace{2 mm} The instrumental resolution for the Gemini spectra was derived as the range of Gaussian FWHMs measured for a 
number of unblended arc lines. The resolution of the TNG spectra have been estimated using the FWHM of the night sky lines.

\end{footnotesize}
\label{log_NIR_spectra}
\end{table*}

\end{appendix}

\end{document}